\documentclass[a4paper,fleqn,usenatbib,useAMS]{mnras}

\usepackage{graphicx}	
\usepackage{amsmath}	
\usepackage{amssymb}	
\usepackage{multicol}        
\usepackage{bm}		
\usepackage{pdflscape}	

\newcommand{\kms}{\,km\,s$^{-1}$} 

\usepackage{gensymb} 
\usepackage{soul} 

\newcommand{\feh}{{\rm {[Fe/H]}}}

\def\afe{{\rm [\alpha/Fe]}}

\def\kms{\,{\rm km~s^{-1}}}
\def\kpc{\,{\rm kpc}}

\def\dex{\,{\rm dex}}

\def\ltsima{$\; \buildrel < \over \sim \;$}
\def\simlt{\lower.5ex\hbox{\ltsima}}
\def\gtsima{$\; \buildrel > \over \sim \;$}
\def\simgt{\lower.5ex\hbox{\gtsima}}

\def\ltsima{$\; \buildrel < \over \sim \;$}
\def\simlt{\lower.5ex\hbox{\ltsima}}
\def\gtsima{$\; \buildrel > \over \sim \;$}
\def\simgt{\lower.5ex\hbox{\gtsima}}

\title[Chemical separation of disc components using RAVE]{Chemical separation of disc components using RAVE}

\author[J. Wojno et al.]{
Jennifer Wojno,$^{1}$\thanks{E-mail: jwojno@aip.de (JW)}
Georges Kordopatis,$^{1}$
Matthias Steinmetz,$^{1}$
Paul McMillan,$^{2}$
\newauthor
Gal~Matijevi\v c,$^{1}$
James Binney,$^{3}$
Rosemary F.\ G.\ Wyse,$^{4}$
Corrado Boeche,$^{5}$ 
\newauthor
Andreas Just,$^{5}$
Eva K.\ Grebel,$^{5}$
Arnaud Siebert,$^{6}$
Olivier Bienaym\'{e},$^{6}$
\newauthor
Brad K. Gibson,$^{7}$
Toma\v z Zwitter,$^{8}$
Joss Bland-Hawthorn,$^{9}$
Julio F.\ Navarro,$^{10}$
\newauthor
Quentin A. Parker,$^{11}$
Warren Reid,$^{12,13}$
George Seabroke,$^{14}$
Fred Watson$^{15}$
\\
$^{1}$Leibniz Institut f\"ur Astrophysik Potsdam, An der Sternwarte 16, 14482 Potsdam, Germany\\
$^{2}$Lund Observatory, Lund University, Department of Astronomy and Theoretical Physics, Box
43, SE-22100 Lund, Sweden \\
$^{3}$Rudolf Peierls Centre for Theoretical Physics, Keble Road, Oxford OX1 3NP, UK \\
$^{4}$Department of Physics and Astronomy, Johns Hopkins University, 3400 N. Charles St, Baltimore, MD 21218, USA \\
$^{5}$Astronomisches Rechen-Institut, Zentrum f\"ur Astronomie der Universit\"at Heidelberg, M\"onchhofstr.\ 12--14, 69120 Heidelberg, Germany \\
$^{6}$Observatoire astronomique de Strasbourg, Universit\'{e} de Strasbourg, CNRS, UMR 7550, 11 rue de l'Universit\'{e}, F-67000 Strasbourg, France \\
$^{7}$E.A. Milne Centre for Astrophysics, University of Hull, Hull, HU6 7RX, United Kingdom \\
$^{8}$Faculty of Mathematics and Physics, University of Ljubljana, 1000 Ljubljana, Slovenia \\
$^{9}$Sydney Institute for Astronomy, School of Physics A28, University of Sydney, NSW 2006, Australia \\
$^{10}$Senior CIfAR Fellow. Department of Physics and Astronomy, University of Victoria, Victoria, BC V8P 5C2, Canada\\
$^{11}$Department of Physics, Chong Yuet Ming Physics Building, The University of Hong Kong, Hong Kong\\
$^{12}$Department of Physics and Astronomy, Macquarie University, Sydney, NSW 2109, Australia \\
$^{13}$Western Sydney University, Locked Bag 1797, Penrith South DC, NSW 1797, Australia \\
$^{14}$Mullard Space Science Laboratory, University College London, Holmbury St Mary, Dorking, RH5 6NT, UK\\
$^{15}$Australian Astronomical Observatory, North Ryde, NSW 2113, Australia \\
}

\date{Accepted XXX. Received YYY; in original form ZZZ}

\pubyear{2016}

\begin{document}
\label{firstpage}
\pagerange{\pageref{firstpage}--\pageref{lastpage}}
\maketitle

\begin{abstract}

We present evidence from the RAdial Velocity Experiment (RAVE) survey of
chemically separated, kinematically distinct disc components in the solar
neighbourhood. We apply probabilistic chemical selection criteria to 
separate
our sample into $\alpha$-low (`thin disc') and $\alpha$-high (`thick disc')
sequences. Using newly derived distances, which will be utilized in the
upcoming RAVE DR5, we explore the kinematic trends as a function of
metallicity for each of the disc components. For our $\alpha$-low
disc, we find a negative trend in the mean rotational velocity
($V_{\mathrm{\phi}}$) as a function of iron abundance ([Fe/H]). We measure a
positive gradient $\partial V_{\mathrm{\phi}}$/$\partial$[Fe/H] for the
{$\alpha$-high disc}, consistent with results from high-resolution
surveys. We also find differences between the $\alpha$-low and
$\alpha$-high
discs in all three components of velocity dispersion. We discuss the
implications of an $\alpha$-low, metal-rich population originating from the
inner Galaxy, where the orbits of these stars have been significantly altered
by radial mixing mechanisms in order to bring them into the solar
neighbourhood. {The probabilistic separation we propose can be
extended to other data sets for which the accuracy in [$\alpha$/Fe] is not
sufficient to disentangle the chemical disc components a priori. For such
datasets which will also have significant overlap with Gaia DR1, we can
therefore make full use of the improved parallax and proper motion data as it
becomes available to investigate kinematic trends in these chemical disc
components.}

\end{abstract}




\section{Introduction}

%

In recent years the study of the history of our Galaxy through detailed
observations of stellar populations has developed into the field known as
Galactic archaeology. From our position within the Milky Way, we have the
unique opportunity to study stellar dynamics and chemistry in great detail.
Large-scale spectroscopic surveys such as RAVE
\citep{Steinmetz06}, SEGUE \citep{Yanny09}, APOGEE \citep{Majewski15},
Gaia-ESO \citep{Gilmore12}, LAMOST \citep{Zhao12}, and GALAH
\citep{DeSilva15} now make it possible to disentangle the
history of star formation and chemical enrichment, and thus to reconstruct
the development of the Galaxy as a whole.

Stars hold chemical information about their birth environment in the
composition of their atmospheres, which remain relatively constant over their
main-sequence lifetime \citep{Freeman02}. If stars remained at their birth
radii throughout their entire lives, and the metallicity of the ISM
increased monotonically, we would expect to observe a tight
correlation between stellar metallicity and age; however, in the
solar neighbourhood, a range of metallicities  has been
observed at a given age \citep[e.g.][]{Edvardsson93,Haywood08,Bergemann14}. For the
oldest stars ($\tau>8\,$Gyr) a correlation between age and metallicity
\emph{is} observed \citep{Haywood13}, and a variety of mechanisms
have been proposed to reconcile the lack of correlation for stars younger
than $8\,$Gyr. The mechanisms include a non-monotonic increase in metallicity
\citep[e.g. inhomogeneities in the early turbulent interstellar medium
(ISM)][]{Haywood13,Bournaud09,Brook04}, and dynamics of stars such that they
are sometimes observed far from their birth radii.

\citet{Sellwood02} discussed the impact of orbital dynamics on
age-metallicity relations in terms of two processes.  As a star ages, the
eccentricity of its orbit increases, widening the radial band within which the star
may be observed. This process they called ``blurring''.  \citet{Sellwood02}
showed that from time to time a star's guiding centre can shift fairly
abruptly to a smaller or larger radius following an interaction with a
transient non-axisymmetric perturbation of the disc. They dubbed this process
``churning''.  Whereas blurring on its own is not powerful enough to account
for the wide range of metallicities present near the Sun at a given age,
\citet{Schoenrich09} argued that churning and blurring working together
account rather nicely for the data from the Geneva-Copenhagen Survey
\citep{Nordstroem04,Holmberg07}.

Of particular interest in the context of radial mixing are stars
with super-solar metallicity, or super metal-rich (SMR)\footnote{For our sample, we define SMR stars as
those with [Fe/H]~$\ga~0.15$.} stars
\citep{Kordopatis15_rich}.  The presence of these metal-rich stars in
the solar neighbourhood has long been problematic for the theory of Galactic
chemodynamics \citep[e.g.][]{Grenon72, Israelian08}. The ISM in the solar
neighbourhood is expected now to be as metal-rich as it has ever been, and is now 
around solar metallicity and relatively homogeneous
\citep{Cartledge06}. If we assume a monotonic metallicity gradient in
the disc, for
SMR stars we infer birth radii $R\la3\kpc$ \citep{Kordopatis15_rich}. 

For lower metallicities, we consider the chemodynamical history of the
thick disc.
While the mechanisms by which
the thick disc formed are hotly debated, the thick disc consists mostly of
old, metal-poor ($\sim -1.5 <$ [Fe/H] $<$ 0.1), $\alpha$-enhanced stars on
kinematically hotter orbits than thin-disc stars \citep{Chiba00, Bensby03}.
Although \citet{Bovy12}, using low-resolution spectra, argued that 
the thin and the thick discs blend continuously into one another, a
trough in the density of stars in the space of $\alpha$-abundance versus
metallicity 
leads the majority of authors to suppose that the thin and thick discs have
experienced different evolutionary histories \citep[e.g.][]{Lee11,Fuhrmann11,Adibekyan13,Haywood13,Bensby14,Recio-Blanco14,Guiglion15,Hayden15,Kordopatis15_gaiaeso}. 
Indeed, \citet{AumerBS} show
that in N-body models scattering of stars by the inevitable non-axisymmetric
features in a disc generates structures remarkably like the thin disc, but do
not generate significant thick discs \citep[see also][but also \citealt{Schoenrich09_rm, Schoenrich09, Loebman11} for how thick discs may be formed due to radial migration]{Minchev12}. They conclude that the thick disc was
present before the thin disc started to form.  By assigning stars as
belonging to either the thin or thick disc according to their chemical
properties, we can explore the possible differences in the chemodynamical
properties of these populations.

In this paper, we aim to identify chemically distinct thin and thick disc
components using the medium resolution ($R \sim 7500$) RAdial Velocity
Experiment (RAVE) survey \citep{Steinmetz06}. The
magnitude-limited ($9<I<12$) RAVE survey offers a kinematically unbiased
sample of stars, ideal for investigating stellar dynamics. The fourth data
release (DR4), presented in \citet[][hereafter K13]{Kordopatis13}, provides
radial velocities, stellar parameters, abundance measurements
\citep{Boeche11}, and distance estimates \citep{Binney14} for 425\,561 stars,
making it one of the
largest spectroscopic surveys with unique
statistical strengths.
Combining these radial velocity data and distance estimates with proper
motions, full 6D (position and velocity) information is available for
the majority of stars.

\defcitealias{Kordopatis13}{K13}

The paper is organized as follows: Section \ref{sec:sampleselection} briefly
describes the quality criteria and parameter cuts applied to obtain the final
working sample, improvements on the data set, and methods to derive stellar
kinematic properties used in the analysis. Section
\ref{sec:chemical_separation} describes the method used to disentangle the
chemical disc components. Section \ref{sec:results} presents the 
kinematic trends for each component. We characterize the trends in mean
$V_{\phi}$ velocity for both the $\alpha$-low and $\alpha$-high populations, and
discuss trends observed in the  dispersions of all three velocity
components. We also estimate the scale lengths of our chemically
selected discs. In Section \ref{sec:discussion} we discuss implications for
these findings, and in particular discuss possible origins of our metal-rich,
$\alpha$-low stars. Finally, Section \ref{sec:conclusion} summarises our
results. 

\section{RAVE data sample and kinematics}
\label{sec:sampleselection}

To ensure a high quality sample of stars in the extended solar neighbourhood,
we use a subsample of RAVE DR4 that meets a number of quality criteria.
First, we select stars with signal-to-noise (SNR) per pixel $>$ 80. We remove stars
that emerge from the chemical abundance pipeline \citep{Boeche11} with
$\mathrm{CHISQ\_c} > 2000$, so retaining only stars with a good fit between
template and observed spectra. In addition, we require the quality flag on
the convergence of the DR4 pipeline to be $\mathrm{Algo\_Conv\_K} \neq
1$.\footnote{A flag of 1 indicates that the pipeline failed to converge (for
more details see \citetalias{Kordopatis13}).} Finally, we utilize stellar
classification flags described in \citet{Matijevic12} to exclude a small
fraction of stars with spectra for which the learning grid contains no
template -- e.g. stars with chromospheric emission \citep{Zerjal13}, spectra
with wavelength calibration problems, carbon stars, and binary stars. 

After these quality cuts have been applied, we remove all stars with
line-of-sight distances greater than $1\,$kpc. We focus our investigation on
the kinematics of stars in the extended solar neighbourhood because the
global properties in this domain have been extensively studied, and the
metallicity gradient $\partial\mathrm{[M/H]}/\partial
R\sim-0.06\kms\kpc^{-1}$, \citep[e.g.][]{Genovali14, Boeche13b}, and velocity
gradient $\partial{V_{R,\phi,Z}}/\partial R\sim\pm3\kms\kpc^{-1}$,
\citep[e.g.][]{Siebert11, Monari14} 
are such that changes in mean velocity
and metallicity across this volume are small. In addition, we require that
the total space velocity, $V_{\mathrm{tot}}$, is less than the Galactic
escape speed, $V_{\mathrm{esc}}$, where we adopt the lower limit
$V_{\mathrm{esc}}\ge492\kms$ determined by \citet{Piffl14}. Our
sample is further reduced because not all RAVE stars have
abundance measurements in DR4. After these cuts, we are left with a sample of
20\,751 stars, which is evenly split between 10\,384 dwarfs and 10\,367 giants.

Galactocentric space velocities in a right-handed cylindrical coordinate
system were determined using the equations summarised in Appendix A of
\citet{Williams13}. First, to transform the observed velocities into a
Galactocentric coordinate system, we adopt values for the solar peculiar
motion with respect to the Local Standard of Rest (LSR) of $(U, V,
W)_{\odot}$ = (11.10, 12.24, 7.25)\,km\,s$^{-1}$ from \citet{Schoenrich10}.
In addition, we take the location of the Sun to be $(R_{0},z_0) =
(8.3,0)\,$kpc and the LSR speed to be $V_{LSR} \sim$\,240\,km\,s$^{-1}$
\citep{Schoenrich12}. When calculating the Galactocentric space velocities,
we use radial velocity measurements from DR4, Galactic coordinates $(\ell,b)$ from 2MASS
\citep{Cutri03} and proper motion measurements from
UCAC4~\citep{Zacharias13}\footnote{While DR4 also provides a number of
sources for proper motion measurements, we find no substantial difference in
our results between two of the most recent catalogues (UCAC4 and
PPMXL~\citep{Roeser10}). We chose to use UCAC4 values, as this catalog is
less affected by potential systematic uncertainties \citep{Vickers16}.}.
Distances have been provided using a version of the RAVE distance
pipeline \citep{Binney14} that has  been updated to include
an extended range of Padova isochrones \citep{Bertelli08} down to
$\mathrm{[Fe/H]} = -2.2$, while it was previously limited to $-0.9\dex$.
These improved distances only systematically affect the results for the most
metal-poor stars.

\section{Chemical separation of the disc}

We separate our sample into two populations using a chemical criterion: the
star's position in plane spanned by $\alpha$-abundance ([Mg/Fe]) versus 
iron abundance ([Fe/H]). While the thin (D) and thick disc (TD)
overlap spatially and kinematically, several studies have shown that it is
possible to disentangle the two components in the [$\alpha$/Fe]-[Fe/H] plane.
These studies include surveys of nearby stars
\citep{Fuhrmann98,Fuhrmann04,Fuhrmann08,Fuhrmann11}, {a low-resolution study
of extended solar neighbourhood G-dwarfs \citep{Lee11}}, and a number of
recent high-resolution studies
\citep{Reddy06, Adibekyan13,Haywood13,Bensby14,Recio-Blanco14,Guiglion15,
Hayden15,Kordopatis15_gaiaeso}.  {RAVE DR4 provides abundance measurements
for six elements derived from the RAVE chemical pipeline,
which includes the $\alpha$ elements Mg, Si, and Ti.
\citetalias{Kordopatis13} suggested that Ti is not reliably measured for
dwarfs, and dwarfs make up half our sample, so we use the
Mg abundance measurement only.}

The precision of [Mg/Fe] abundances determined with RAVE ($\sim0.2\,$dex) is
lower than that required to recover the gap between the two populations in
the [$\alpha$/Fe]-[Fe/H] plane: using Eq.~3 of \citet{Lindegren13}, which
describes the minimum separation at which two populations can be
distinguished in a given sample size, we find that with our sample of 20\,751
stars we could distinguish populations separated by 1.5 times the standard
error in [Mg/Fe]. Hence given our error of $\sim0.2\,$dex in [Mg/Fe], the
separation between populations needs to be at least $0.3\,$dex, whereas
high-resolution data indicate that the separation is $\la0.2\,$dex
\citep{Recio-Blanco14,Haywood13,Bensby14,Kordopatis15_gaiaeso}. Therefore, we
turn to a probabilistic approach to the separation of the $\alpha$-low and
$\alpha$-high stars. 

\begin{figure}
	\includegraphics[width=1.05\columnwidth]{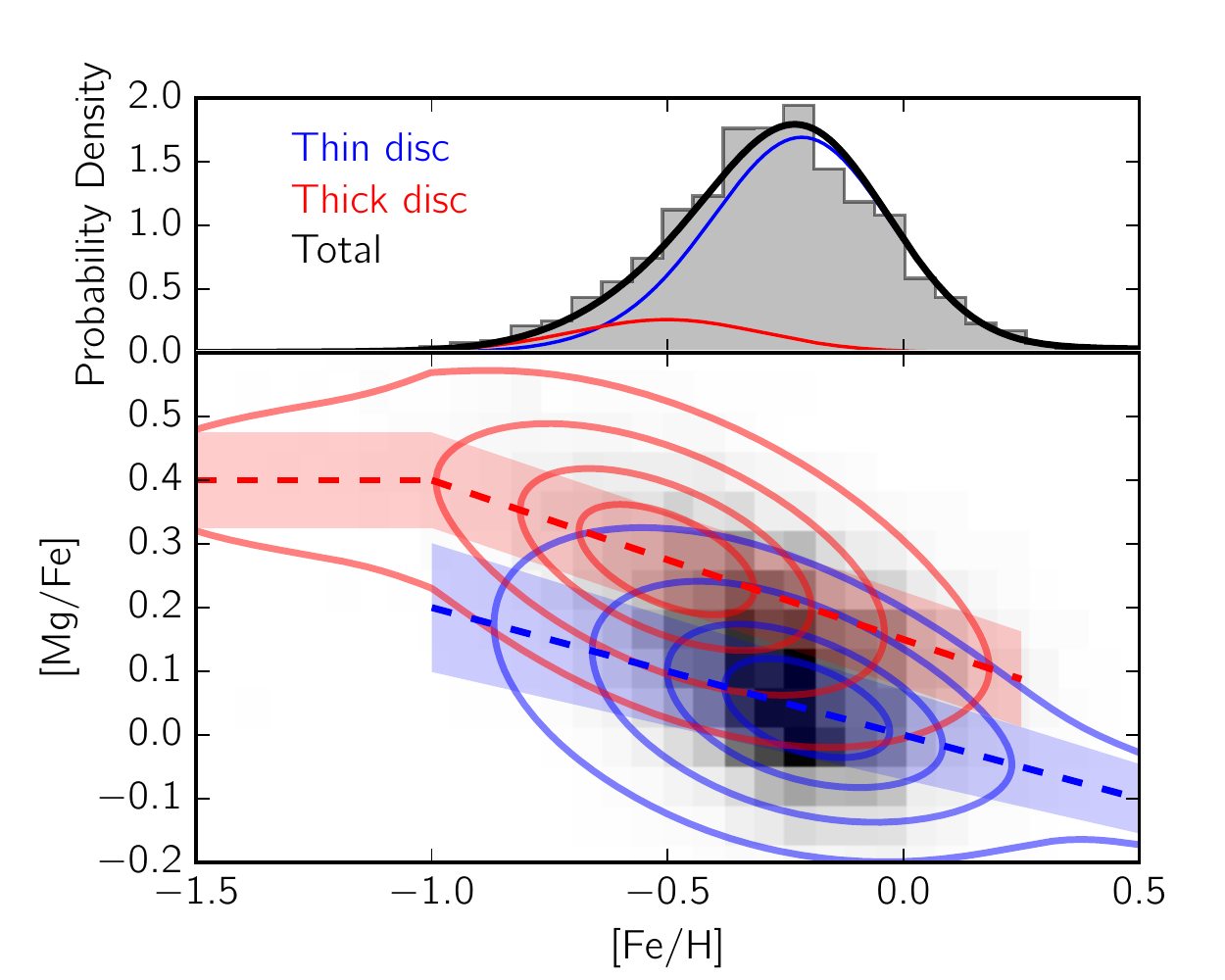}
    \caption{\textit{Top panel}: adopted metallicity distribution function
(MDF). The grey-shaded histogram represents the MDF of our final sample. The
thin and thick disc MDFs are shown in blue and red, respectively, and the
thick black line is the sum of both disc components. \textit{Bottom panel}:
{2D histogram of our sample in [Mg/Fe]-[Fe/H] space.} Contours show
33, 67, 90, and 99\% of thin disc (blue) and thick disc (pink) 2D pdfs. The over-plotted dashed blue line shows the
assumed mean ($\mu_{\mathrm{Mg}}$) variation as a function of [Fe/H], where the
filled area represents the assumed variation in the
$\sigma_{\mathrm{Mg}}$ as a function of [Fe/H], for the thin disc (Eqs.
\ref{eq:mu_alpha_d} and \ref{eq:sigma_alpha_d}). Similarly, the over-plotted
dashed red line shows the adopted thick disc relations (Eqs.
\ref{eq:mu_alpha_td} and \ref{eq:sigma_alpha_td}).}
    \label{fig:mdf}
\end{figure}

We write the 2D probability density function (pdf) of stars in the
$\alpha$-low or the $\alpha$-high component as
 \begin{equation}
f(\feh,\afe) =   f_{\mathrm{[Mg/Fe]}}   \times f_{\mathrm{[Fe/H]}}.
\label{eq:f_tot}
\end{equation}
We will refer to $f_{\mathrm{[Mg/Fe]}}$ as the $\alpha$ distribution
function, and $f_{\mathrm{[Fe/H]}}$ as the metallicity distribution function
(MDF) for a given component.  The $\alpha$ distribution is taken to be a
Gaussian with a mean and dispersion that depends on [Fe/H]: 
\begin{equation}
f_{\mathrm{[Mg/Fe]}} = \frac{1}{\sigma_{\mathrm{Mg}} \sqrt{2 \pi}} 
\mathrm{exp}\left(-\frac{(\mathrm{[Mg/Fe]} - \mu_{\mathrm{Mg}})^{2}}{2
{\sigma_{\mathrm{Mg}}}^2} \right) .
\label{eq:f_alpha}
\end{equation}
For the thin disc,  $\mu_{\mathrm{Mg}}$ and
$\sigma_{\mathrm{Mg}}$ are given by
\citet{Kordopatis15_gaiaeso}:
{\arraycolsep=1pt
\begin{eqnarray}
\mu_{\mathrm{Mg, D}} &=& -0.2 \times \mathrm{[Fe/H]}
 \label{eq:mu_alpha_d} \\
\sigma_{\mathrm{Mg, D}} &=& -0.031\times \mathrm{[Fe/H]} + 0.07 .
\label{eq:sigma_alpha_d} 
\end{eqnarray}
For the thick disc, $\mu_{\mathrm{Mg}}$ and $\sigma_{\mathrm{Mg}}$ 
are given by:
\begin{eqnarray}
\mu_{\mathrm{Mg, TD}} &=&
\begin{cases}0.4 &\hbox{for }\mathrm{[Fe/H]} < -1.0\\
 -0.25\times \mathrm{[Fe/H]} + 0.15&\hbox{otherwise}
\end{cases} \label{eq:mu_alpha_td}\\
\sigma_{\mathrm{Mg, TD}} &=& 0.075 .
\label{eq:sigma_alpha_td}
\end{eqnarray}}
In the lower panel of Fig.~\ref{fig:mdf} the linear dependences of
$\mu_{\mathrm{Mg}}$ on [Fe/H] are shown by the dashed lines (blue for the
$\alpha$-low and red for the $\alpha$-high component), while the widths of
the blue and red shaded regions around these lines indicate the values of
$\sigma_{\mathrm{Mg}}$ for both the $\alpha$-low and -high components.

The MDFs are too skew to be satisfactorily represented by a single Gaussian,
so we represent them as weighted sums of multiple Gaussians,
\begin{equation}
f_{\mathrm{[Fe/H]}} = \sum_{i=1}^{n} \frac{a_{i}}{\sigma_{\mathrm{Fe},i} \sqrt{2 \pi}} \mathrm{exp}\left(-\frac{(\mathrm{[Fe/H]} - \mu_{\mathrm{Fe},i})^{2}}{2 {\sigma_{\mathrm{Fe},i}}^2}\right) ,
\label{eq:f_met}
\end{equation}
{where $n = 3$ for the thin disc and $n = 2$ for the thick disc.} The means, dispersions, and weights $a_i$ of 
these Gaussians for both the
thin and thick disc can be found in Table~\ref{tab:gauss_values}.  These
values were extracted using the high-resolution measurements of
\citet{Kordopatis15_gaiaeso} as a starting point, but the shapes of the
distribution were slightly modified because the shape of the MDF in any
survey is affected by the survey's selection function. The top panel of
Fig.~\ref{fig:mdf} shows the adopted model MDF, using the values from
Table~\ref{tab:gauss_values}. {We do not consider a metal-rich tail for the 
thick disc due to ambiguity regarding the mixture of populations in the $\alpha$-high, metal-rich region of the $\afe - 
\feh$ plane (more details regarding potential complications due to $\alpha$-high, metal-rich stars can be found 
later in this section as well as in Sec.~\ref{sec:conclusion}).} 

\begin{table}
	\centering
	\caption{Parameters for Eq. \ref{eq:f_met} for both thin (D) and
thick (TD) disc components.}
	\begin{tabular}{lcr}
		\hline
		 & D & TD \\
		\hline
		$a_{1}$ & 0.8 & 0.9\\
		$\mu_{\rm Fe,1}$ & -0.2 & -0.5\\
		$\sigma_{\rm Fe,1}$ & 0.18 & 0.2\\
		
		$a_{2}$ & 0.15 & 0.08\\
		$\mu_{\rm Fe,2}$ & -0.4 & -0.8\\
		$\sigma_{\rm Fe,2}$ & 0.2 & 0.4\\
		
		$a_{3}$ & 0.05 & --\\
		$\mu_{\rm Fe,3}$ & 0.2 & --\\
		$\sigma_{\rm Fe,3}$ & 0.5 & --\\
		$X$ & 0.85 & 0.14\\
		\hline
	\end{tabular}
	\label{tab:gauss_values}
\end{table}

\label{sec:chemical_separation}

\begin{figure}
	\includegraphics[width=\columnwidth]{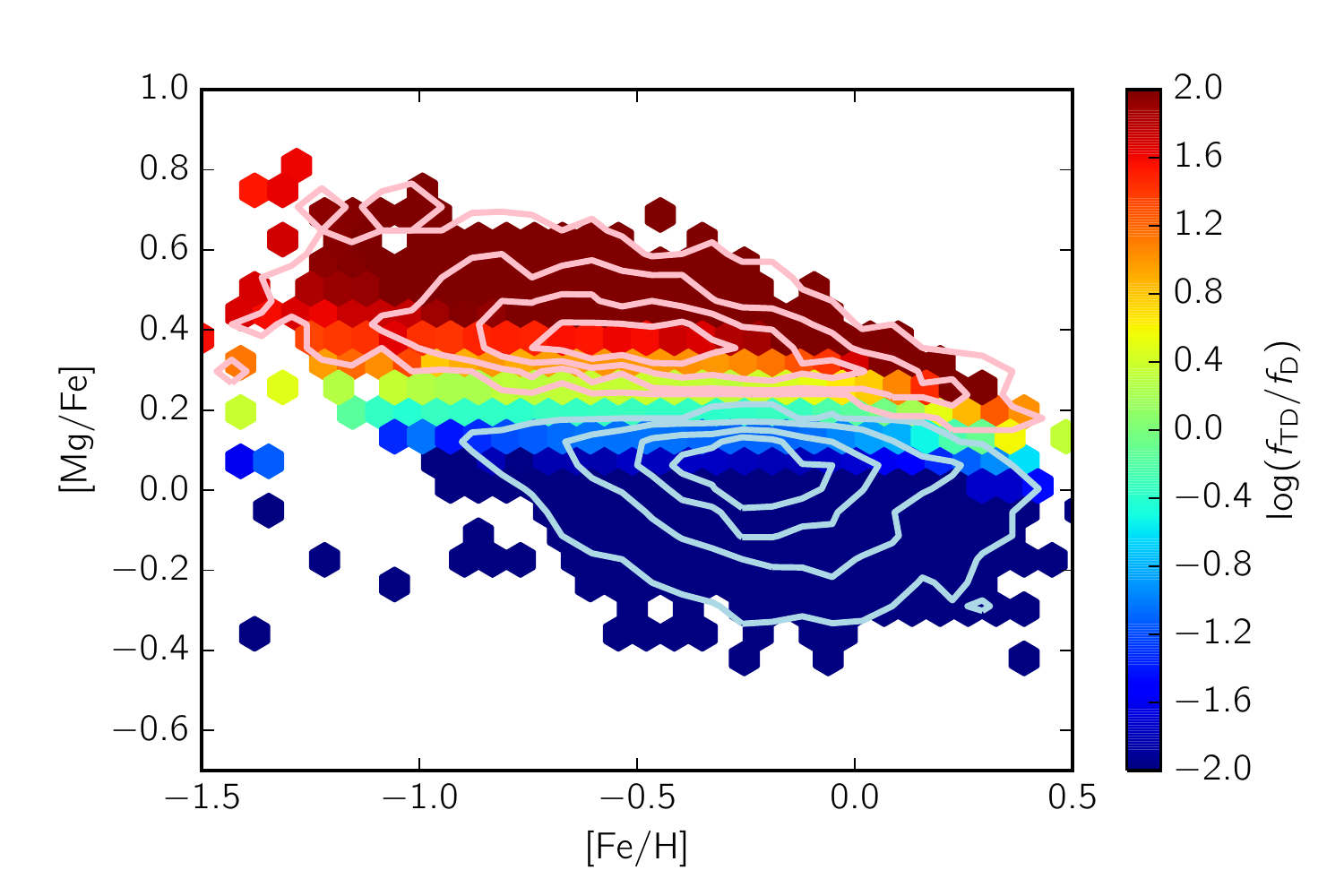}
    \caption{[Mg/Fe] v. [Fe/H], where the colour of the bins represent the
logarithm log$_{10}$($f_{\mathrm{TD}}/f_{\mathrm{D}}$) of the average
relative probability of stars in that bin belonging to either the thick
or the thin disc. Stars with a relative probability between 0.1 and
10 are not subsequently used but are included here to demonstrate the gradient in
probability. The contours show 33, 67, 90, and 99\% of the selected thin
disc (blue) and thick disc (pink) distributions, to illustrate the gap
between the two components.}
    \label{fig:alpha_met}
\end{figure}

The pdfs given by Eq.~\eqref{eq:f_tot} are normalised to unity.
{Since there are believed to be more stars in the thin disc than the thick
disc \citep{Bland-Hawthorn16}, we multiply these normalised pdfs by factors $X_{\mathrm{TD}}$ and
$X_{\mathrm{D}}$ equal to the probabilities that a randomly chosen star in
the sample belongs to the thick or thin disc.  The values of
$X_{\mathrm{TD}}$ and $X_{\mathrm{D}}$ are not accurately known.}
{RAVE is kinematically unbiased, and as a magnitude-limited survey
that extends much further than one thin-disc scale height, we expect
thin and thick disc stars to enter the survey roughly in proportion to the
local surface densities of the two discs. Our adopted values are based on the
results of \citet{Soubiran03}, who separated the thin and thick disc
populations in velocity-metallicity space.  They give an estimate of
$15\pm7\%$ for the local normalization of the thick disc and $85\pm7\%$ for the thin disc, where the median distance of their sample is 400~pc from the Galactic plane.  The median distance of our sample is similar, so we find it appropriate to adopt these values, modifying them slightly.}  We use $X_{\mathrm{TD}}=0.14$ and
$X_{\mathrm{D}}=0.85$ (see Table~\ref{tab:gauss_values})\footnote{{As these values are widely disputed,
we explore the effects of removing the $X_{\mathrm{TD}}$ and $X_{\mathrm{D}}$ factors from our probability 
computations (i.e., giving both disc 
components equal weight). The corresponding results can be found in Sections~\ref{sec:vel_means} and \ref{sec:vel_dispersion}.}}. Thus we assume that $\sim 1$\% of the stars in the
extended solar neighbourhood belong to the halo. We do not explicitly
calculate the probability of a star belonging to the halo, however.  {\citet{Chen01} note the degeneracy between local normalization and surface density ratio due to the uncertain nature of the scale heights of the discs. However, our chosen value is also a conservative estimate when considering the 
overall relative thin/thick disc surface density ratio \citep[see][]{Bland-Hawthorn16}.}

\begin{figure*}
	\includegraphics[width=\textwidth]{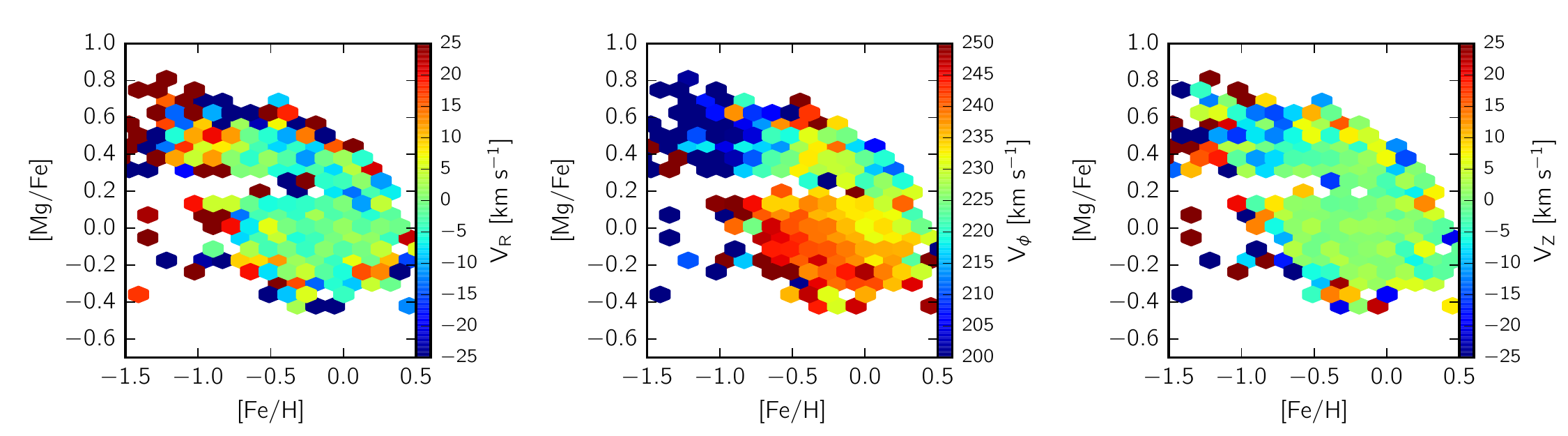}
    \caption{{[Mg/Fe] vs. [Fe/H], where the colour of the bins represent the
average velocity of the stars in that bin, for each of the three velocity
components. Here, we show only the selected stars, i.e., all stars with $0.1
< (f_{\mathrm{TD}}/f_{\mathrm{D}}) < 10$ have been removed.}}
    \label{fig:alpha_met_vel}
\end{figure*}

\begin{figure*}
	\includegraphics[width=\textwidth]{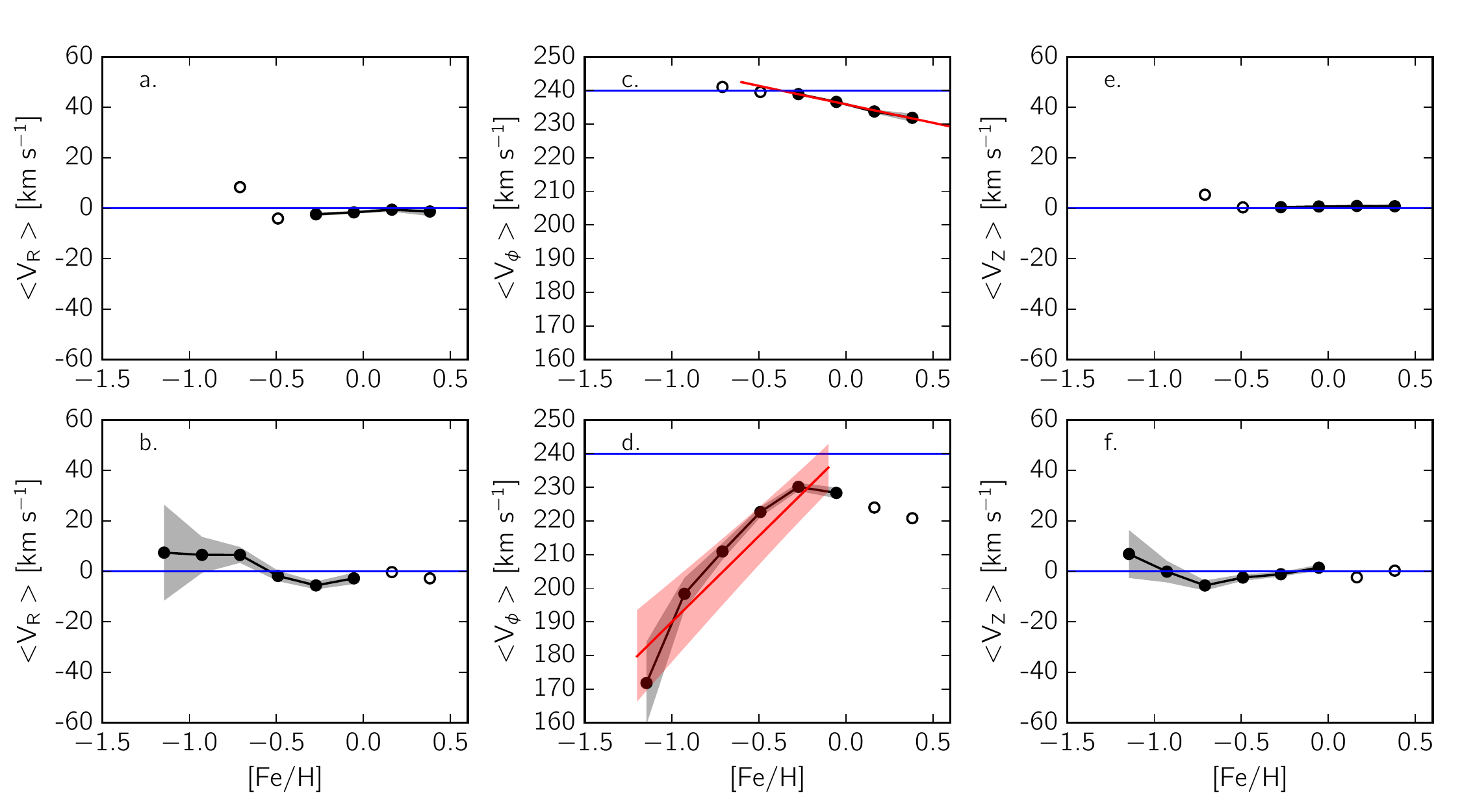}
    \caption{Mean velocity as a function of [Fe/H] for the {$\alpha$-low} component (top row), and the {$\alpha$-high} component (bottom row). See Section~\ref{sec:chemical_separation} for a detailed description on how these populations are selected. Trend lines are computed by binning the data into $\sim$0.2 dex wide [Fe/H] bins. The shaded regions correspond to the average errors for a given metallicity bin. Bins with less than 10 stars are not shown, and are not used to calculate the linear fit. Open circles denote bins which are not used in determining the linear fit in panels c and d. The red lines in panels c and d show the linear fits for the {$\alpha$-low and -high sequences}, respectively, with the shaded red region corresponding to the error on the fit.}
    \label{fig:vphi_met}
\end{figure*}


In order to include realistic errors for our [Fe/H] and [Mg/Fe] measurements
(of the order of 0.17 dex and 0.2 dex, respectively), we generate 500
realizations of each star assuming a Gaussian error distribution for both
measurements. The ratio of the thick disc probability to the thin disc
probability $(f_{\mathrm{TD}}/f_{\mathrm{D}})$ is calculated for each
realization, and the median is taken as the final
$f_{\mathrm{TD}}/f_{\mathrm{D}}$ value. We assign a star as belonging to the
thin disc when $f_{\mathrm{TD}}/f_{\mathrm{D}} < 0.1$. Similarly, we assign a
star as belonging to the thick disc when $f_{\mathrm{TD}}/f_{\mathrm{D}} >
10$. {We note that we still expect some overall contamination of incorrectly
assigned stars in both disc components, of the order of 10\%, due to our
selection criterion ($f_{\mathrm{TD}}/f_{\mathrm{D}}$), and this contamination
increases towards both ends of our metallicity distribution. This
contamination may have a number of sources. Stars may be incorrectly assigned
to the thick disc due to the substantial overlap in the MDF of the thin and
thick disc at the metal-rich end. In addition, the precision of our abundance
measurements may affect the assignment accuracy at both tails of the
distribution.}

Despite this contamination, we consider that the $\alpha$-low component corresponds to what would typically be described as the `thin disc', and the $\alpha$-high similarly corresponds to the `thick disc'. In the remainder, we refer to the `thin disc' as the $\alpha$-low
component and the `thick disc' as the $\alpha$-high component. We adopt
this definition due to certain limitations when considering the tails of
[Fe/H] distribution for both the `thin' and 'thick' disc components, where
more complex population mixtures may exist
\citep{Nissen10,Masseron15,Chiappini15,Martig15}. {Specifically, for the metal-rich tail of the $\alpha$-high 	
component, it is unclear if the distribution flattens \citep[e.g.][]{Bensby14}, or continues with a linear trend 
similar to our adopted distribution, eventually joining the $\alpha$-low sequence. In either case, disentangling  
these complex population mixtures in the metal-rich tail is not possible when considering the precision of our $
\alpha$-abundance measurements, and therefore we consider only a 
conservative range of metallicities where a two-component model is viable.}

Our final sample consists of 11\,440 stars assigned to the
$\alpha$-low component and 2\,251 stars assigned to the $\alpha$-high
component. In Fig. \ref{fig:alpha_met}, we show [Mg/Fe] versus [Fe/H], with
bins colour-coded according to the average probability of stars in that bin
to belong to either component.
The contours enclose 33, 67, 90, and 99\% of the selected $\alpha$-low (blue) and $\alpha$-high (pink) components, to illustrate the gap
between the two components.

\section{Results}
\label{sec:results}

\begin{figure}
\begin{center}
	\includegraphics[width=0.75\columnwidth]{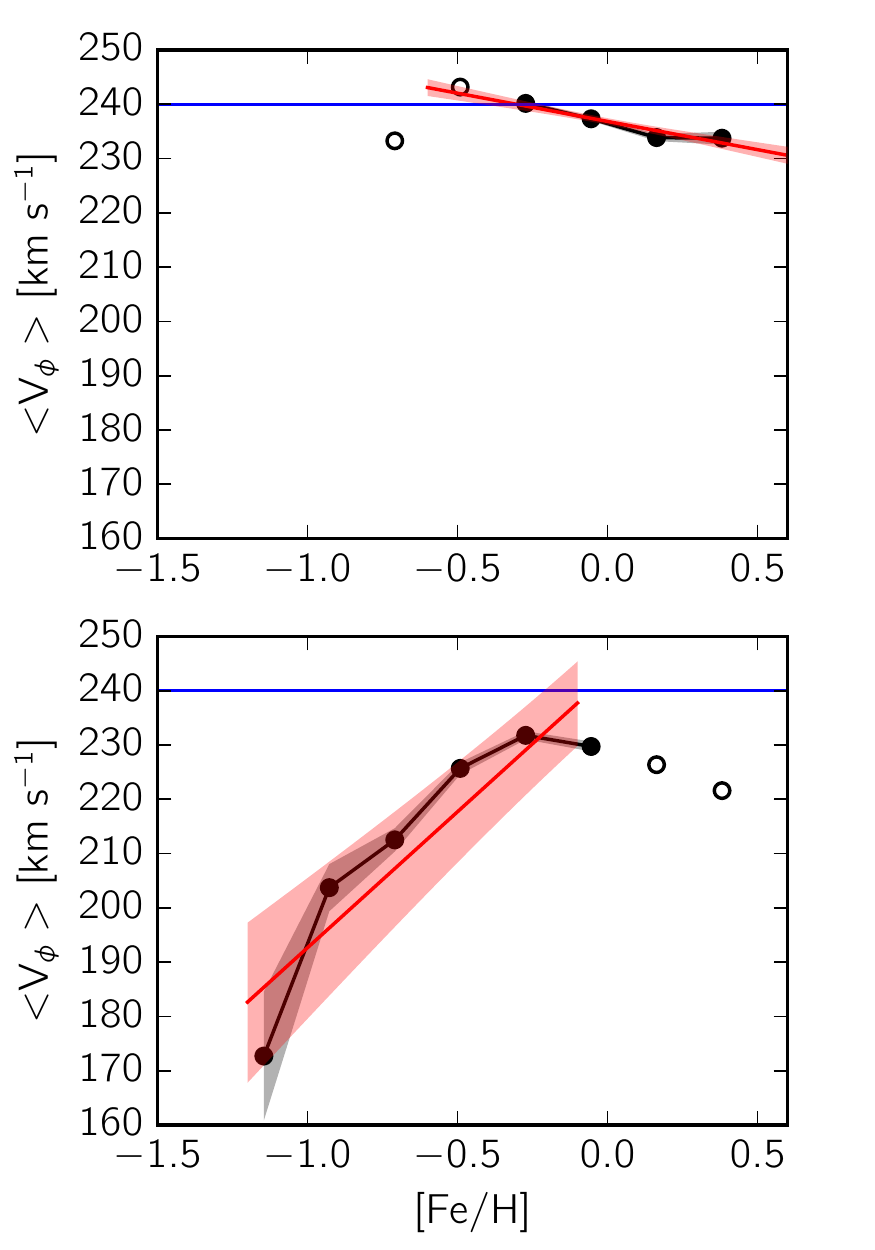}
    \caption{Same as panels c and d of Fig.~\ref{fig:vphi_met}, but with no prior factors used in the determination of the probabilities (i.e., equal weight given to both disc components, cf. to the $X_{\mathrm{TD}}$ and $X_{\mathrm{D}}$ factors given in Table~\ref{tab:gauss_values}).}
    \label{fig:vphi_met_noprior}
\end{center}
\end{figure}

Each pixel of the $\feh - [\mathrm{Mg}/\mathrm{Fe}]$ planes of Fig.~\ref{fig:alpha_met_vel} is
coloured to encode the mean value of $V_R$, $V_\phi$ or $V_Z$ for the stars
in our selection that are assigned to that pixel. A clear distinction is
evident between the kinematics of our selected $\alpha$-low and
$\alpha$-high components, especially in $V_\phi$. 

To explore further the different kinematic trends in our chemical components,
in Fig.~\ref{fig:vphi_met} we show for each component the averages of $V_R$,
$V_\phi$ and $V_z$ as a function of metallicity. In the case of $V_\phi$ we
have fitted by least-squares lines to the data points, excluding bins with
less than ten objects to avoid drawing conclusions from low-number
statistics.  For the {$\alpha$-low sequence}, the most metal-poor bins
are also excluded as there we expect some contamination from the
{$\alpha$-high component} (see Sec.~\ref{sec:chemical_separation}).
For the $\alpha$-high disc (bottom row of Fig.~\ref{fig:vphi_met}), we
exclude the two most metal-rich bins, due to contamination from $\alpha$-low
stars. Bins excluded from the linear fits are
plotted as open circles.

\subsection{Mean rotational velocity trends for the thin and thick disc components} 
\label{sec:vel_means}

\begin{figure*}
	\includegraphics[width=\textwidth]{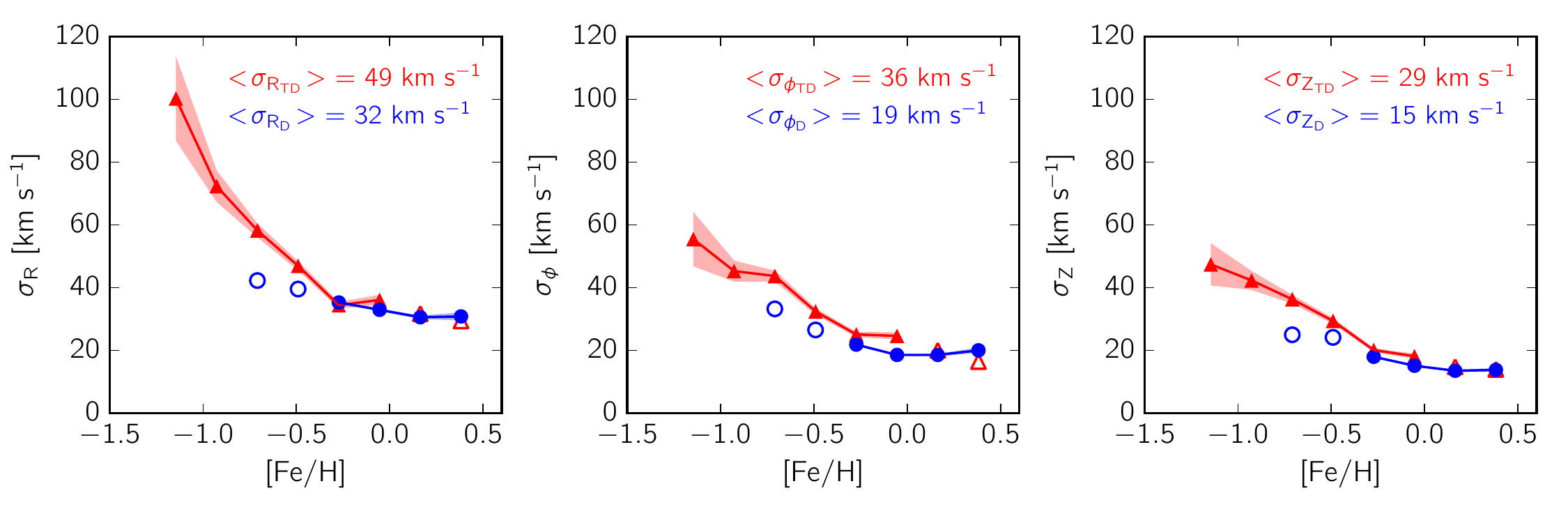}
    \caption{Velocity dispersion as a function of metallicity for the {$\alpha$-low} component (blue circles), and the {$\alpha$-high} component (red triangles). See Section~\ref{sec:chemical_separation} for a detailed description on how these populations are selected. Trend lines are computed by binning the data into $\sim$0.2 dex width bins. The shaded regions correspond to average errors for a given metallicity bin. Bins with less than 10 stars are not shown. The average $\sigma_{\mathrm{R},\phi,\mathrm{Z}}$ values are given in the top right corner, for both {disc components}. These averages are determined using only bins with filled symbols.}
    \label{fig:vphi_met_std}
\end{figure*}

\begin{figure*}
	\includegraphics[width=\textwidth]{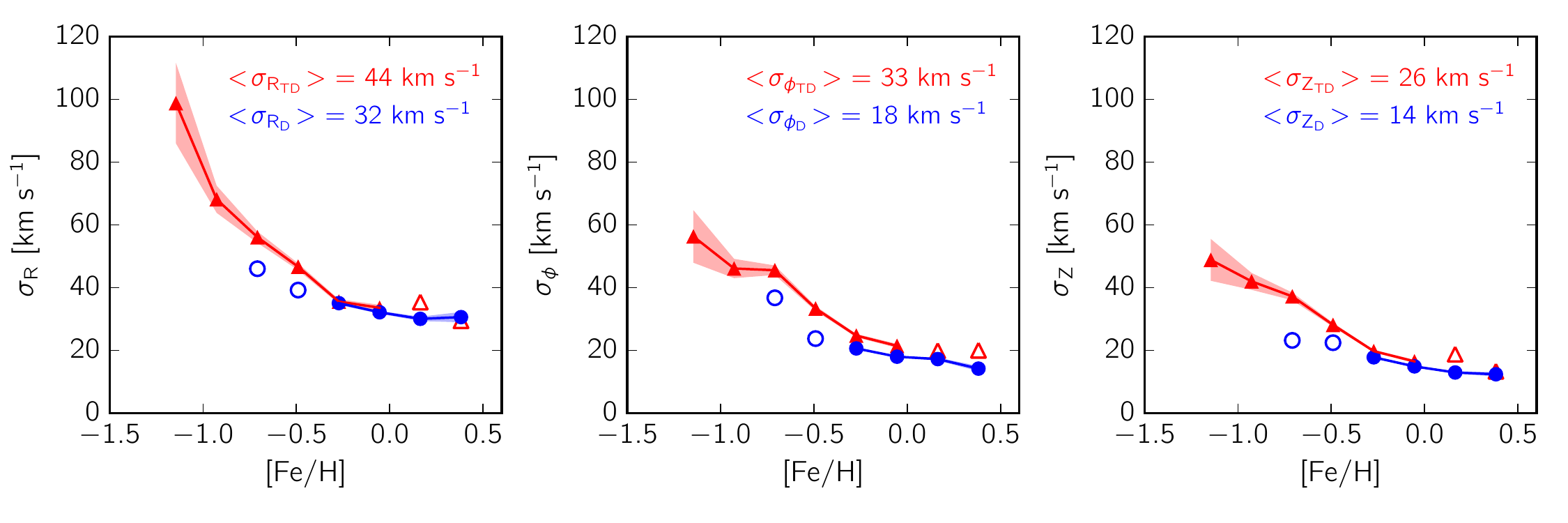}
    \caption{Same as Fig.~\ref{fig:vphi_met_std}, but with no prior factors used in the determination of the probabilities (i.e., equal weight given to both disc components).}
    \label{fig:vphi_met_std_noprior}
\end{figure*}

We find   $\partial V_{\phi}/\partial\feh<0$ for the 
{$\alpha$-low component} but  $\partial V_{\phi}/\partial\feh>0$ for the 
{$\alpha$-high component}. 
Quantitatively, for {$\alpha$-low} stars, we find
\begin{equation}
{\partial V_{\phi}\over
\partial\mathrm{[Fe/H]}}=(-11\pm1)\kms\dex^{-1}.
\end{equation}
This slope is significantly shallower than those of \citet{Lee11} and
\citet{Adibekyan13}, but using high-resolution data from \textit{Gaia}-ESO 
 \citet{Recio-Blanco14} find, $\partial V_{\phi}/\partial\mathrm{[M/H]} =
(-17\pm6)\kms\dex^{-1}$, which lies within 1$\sigma$ of our value.

For the {$\alpha$-high} population, we measure 
\begin{equation}
{\partial
V_{\phi}\over\partial\mathrm{[Fe/H]}} = (51\pm10)\kms\dex^{-1}.
\end{equation}
 This slope is in agreement with recent literature values. \citet{Lee11}
report a tight correlation with slope $(43.4\pm1.8)\kms\dex^{-1}$ in their
subsample of thick-disc G-dwarfs, while \citet{Kordopatis11} find $\partial
V_{\phi}/\partial\mathrm{[M/H]} = (43\pm11)\kms\dex^{-1}$.  Similarly,
\citet{Adibekyan13} find $\partial
V_{\phi}$/$\partial\mathrm{[M/H]}\sim42\kms\dex^{-1}$ for thick disc stars 
in a sample of FGK solar neighbourhood
dwarfs. Our value also agrees with the findings of \citet{Recio-Blanco14},
$\partial V_{\phi}$/$\partial\mathrm{[M/H]} = (43\pm13)\kms\dex^{-1}$, from a
high-resolution sample of FGK stars from the \textit{Gaia}-ESO survey.

{Fig.~\ref{fig:vphi_met_noprior} shows the average $V_\phi$ velocity as a function of 
[Fe/H] for the case where both discs are given equal weight (compare with panels c and d of Fig.~
\ref{fig:vphi_met}). We measure similar trends in $\partial V_{\phi}/\partial\mathrm{[Fe/H]}$ as when the  $X_{\mathrm{TD}}$/$X_{\mathrm{D}}$ prior is included, for both disc components. This further strengthens our result that we successfully determine distinct kinematics for each disc component.}

\subsection{Velocity dispersion trends}
\label{sec:vel_dispersion}

Fig. \ref{fig:vphi_met_std} shows for the $\alpha$-low and $\alpha$-high
components the variation with [Fe/H] of the dispersions of $V_R$, $V_\phi$
and $V_z$ corrected for observational uncertainties in the standard way:
\begin{equation}
\sigma_{\mathrm{R},\phi,\mathrm{Z}} = \sqrt{\sigma_{\mathrm{R},\phi,\mathrm{Z}}^{* 2} - \langle\mathrm{e}_{V_{\mathrm{R},\phi,\mathrm{Z}}}\rangle^{2}}  
\end{equation}
where $\sigma_{\mathrm{R},\phi,\mathrm{Z}}$ is our corrected velocity dispersion, $\sigma_{\mathrm{R},\phi,\mathrm{Z}}^{*}$ is the measured velocity dispersion, and $\mathrm{e}_{V_{\mathrm{R},\phi,\mathrm{Z}}}$ is the error on the velocity component. 
In both chemical subgroups velocity dispersion tends to decrease with
{increasing [Fe/H]}, so the most metal-poor stars have the highest
dispersions and SMR stars have low dispersions. In the region $-0.2 <
\mathrm{[Fe/H]} < 0.0$ the trends for the $\alpha$-low and
$\alpha$-high components overlap substantially: their dispersions differ
little in the two overlapping bins where both the $\alpha$-low and
$\alpha$-high components are free of significant contamination. However, at $\mathrm{[Fe/H]}< -0.2$
the values for each chemical subgroup are clearly separated. While we assume
some contamination of the $\alpha$-low sample in the most metal-poor
bins, these bins follow approximately the same linear relation as the
more metal-rich bins. We find the ratio of $\sigma_{R}$/$\sigma_{\phi}$ to be
relatively constant ($\sim 1.6$ for both components), independent of
metallicity.

At the top right of each panel in Fig.~\ref{fig:vphi_met_std} we give the
mean velocity dispersion for each chemical component, calculated using only
the bins that we consider free of significant contamination (i.e., avoiding
the two most metal-poor bins for the {$\alpha$-low component}, and the
two most metal-rich bins for the {$\alpha$-high component}). This
gives an effective range of $-0.27<\mathrm{[Fe/H]}<0.38$ for the
{$\alpha$-low component}, and an effective range of $-1.15 <
\mathrm{[Fe/H]} < -0.05$ for the {$\alpha$-high} component.

For both {components}, our mean velocity dispersions conform to the
familiar relations $\sigma_{{R}} > \sigma_{\phi} > \sigma_{{Z}}$ and
$\sigma_{{Z}} \simeq 0.5\sigma_{{R}}$ \citep{Quillen01,Holmberg07}. In
addition, we find a constant offset $\sim 16\kms$ between the average
dispersions of the {$\alpha$-low and $\alpha$-high sequences}. This
offset is consistent with the observed values of \citet{Bensby05}, which were
determined using a kinematically-selected high-resolution sample of FGK
dwarfs in the solar neighbourhood. {As in Section~\ref{sec:vel_means}, we consider the velocity
dispersion as a function of [Fe/H] for the case where we remove the $X_{\mathrm{TD}}$/$X_{\mathrm{D}}$  prior from the 
probability calculation. The results are shown in  Fig.~\ref{fig:vphi_met_std_noprior}. While the outcome 
is similar, we do note that the thick disc has a consistently lower mean velocity dispersion compared to the case
where we include the prior. In addition, the separation between the two chemical disc components is also consistently less ($\sim 13 \kms$ compared to when the prior is included $\sim 16\kms$, cf. Figs.~\ref{fig:vphi_met_std} and \ref{fig:vphi_met_std_noprior}). }

{While the removal of the $X_{\mathrm{TD}}$/$X_{\mathrm{D}}$ factor does not significantly alter our results, the ratio of 
stars assigned to the $\alpha$-high and -low components is affected adversely. The ratio of $\alpha$-high to $
\alpha$-low stars for our final selected sample is $2251/11440 \sim 0.20$, which is within the range of 
conservative estimates of 
the thick disc fraction. In contrast, if we remove the prior, we find a ratio of $4436/7785 \sim 0.60$, which is 
above the high end of the range of estimates. In order to better represent physical reality as it is currently 
understood, we find the application of this prior desirable.}

Combining the mean and dispersion of each of velocity component with the
Jeans equation, we can estimate the radial scale lengths ($h_{{R}}$)
of our chemical {disc components}. If we assume that the Galactic potential
is dominated by a centrally concentrated mass distribution \citep{Gilmore89},
and that the velocity ellipsoid always points toward the Galactic centre
\citep{Siebert08,Pasetto12,Binney14_dynamics}, we have:
\begin{equation}
h_{{R}} = \frac{2 R \sigma_{{R}}^{2}}{V_{c}^{2} - \langle
V_{\phi}\rangle^{2} + 2\sigma_{{R}}^{2} - \sigma_{\phi}^{2} 
- \sigma_{{Z}}^{2}}  \;\;.
\end{equation}
By this reckoning our {$\alpha$-low component} has scale length
$h_{{R}_{\mathrm{D}}} = 4.8\pm0.2\,$kpc, and our {$\alpha$-high
component} has $h_{\mathrm{R}_{\mathrm{TD}}} = 3.4\pm0.1\,$kpc.  Our
estimates are consistent with the finding of \citet{Bovy12_sl} from a sample
of local dwarfs from SEGUE, that the scale length of the thin disc is more
extended than that of the thick disc. When we consider the scale lengths of the
discs as functions of metallicity, {for the
$\alpha$-low component} $\partial h_R/\partial\feh<0$
(i.e., the most metal-poor bins have the longest scale lengths).  The scale
length of the {$\alpha$-high component} proves relatively constant,
with only a slight negative trend, with increasing metallicity.

While {the scale length of our $\alpha$-low component} is
longer than that determined by {\citet{Robin03} and
\citet{Juric08} from star counts}, it lies within $2\sigma$ of high-end
estimates, such as that, $h_R=(4.3\pm0.2)\,$kpc, of \citet{Bovy12_sl}, who
used a sample of dwarfs with $\feh\sim 0$. {Our value of $h_R$ for the
$\alpha$-high component is also} larger than literature values. Using a
handful of red giant stars \citet{Bensby11} obtain $h_R\sim 2\,$kpc for the
thick disc, and \citet{Bovy15} find $h_R=(2.2\pm0.2)\,$kpc for an
$\alpha$-high population.  However, we note that the values of our scale
lengths depend sensitively on the adopted value for the peculiar motion of
the Sun ($V_{\sun}$). A smaller $V_{\sun}$ (e.g., the classical value of
$5.25\kms$ \citep{Aumer09}) would result in much smaller scale lengths for
both discs. For a detailed discussion see \citet{Golubov13}.

Many of the previous studies cited use either dwarfs or giants for their
samples, with a tendency to select only dwarfs for chemodynamical studies, as
their atmospheres stay relatively constant until they leave the main sequence. Giants, on the other hand, can experience a significant amount of
mixing in their atmospheres, which makes them less desirable for studies
involving long dynamical timescales. As we have an equal number of dwarfs and
giants in our original sample before the chemical disc selection, we
investigated if these observed trends are affected when considering only
dwarfs, or only giants. For both the mean velocity and velocity dispersion
trends, we find no significant differences in our conclusions.

\section{Discussion}
\label{sec:discussion}

The primary difference in the kinematics of our {$\alpha$-low and
$\alpha$-high components} are the trends in mean rotational velocity as a
function of [Fe/H]. We find opposite signs for $\partial
V_{\phi}$/$\partial\mathrm{[Fe/H]}$ for our two chemical disc components:
positive for the {$\alpha$-high component}, and negative for
the {$\alpha$-low component} (see panels c and d of
Fig.~\ref{fig:vphi_met}).

The {$\alpha$-high component} exhibits the expected characteristics of
asymmetric drift: as [Fe/H] and rotational velocity increase, the velocity
dispersion decreases. On account of asymmetric drift, the `thick' disc lags
the LSR significantly -- \citet{Bensby05} find a lag of 46\,km\,s$^{-1}$
for the kinematically selected thick disc. While the value of the thick
disc's lag is uncertain \citep[cf.][]{Chiba00,Fuhrmann04,Lee11}, positive
values for $\partial V_{\phi}$/$\partial\mathrm{[Fe/H]}$ have also been
observed in a number of previous studies
\citep{Lee11,Kordopatis11,Adibekyan13,Recio-Blanco14}. \citet{Recio-Blanco14}
suggest that this positive trend {is in agreement with the scenario
described in \citet{Haywood13} and \citet{Haywood16}: each subsequent stellar
generation in the thick disc is kinematically cooler than the previous one,
such that the dispersion (and therefore lag behind the LSR) decreases with
increasing [Fe/H].} 

{\citet{Haywood13} then propose that the inner
disc is formed with the properties of the most metal-rich thick disc
`layer', with quenching of star formation causing the corresponding gap
found in the [$\alpha$/Fe]-[Fe/H] plane. What could have caused star
formation to pause  at the end of the formation of the thick disc is still
debated, although a few scenarios have been proposed (e.g., the formation of
the bar \citep{Haywood16} and depletion of gas in the disc \citep{Chiappini97}).
Star formation is then assumed to resume, albeit at a lower rate, in the thin
disc \citep{Just10}.}

Our {$\alpha$-low} component also lags the LSR, but with $\partial
V_{\phi}/\partial\mathrm{[Fe/H]}$ negative. {We propose that the SMR
stars ($\feh\ga0.15$) in our $\alpha$-low component have undergone
significant changes in their orbital kinematics.} The origin of these
metal-rich stars is not immediately obvious, but it is likely that they did
not form locally. In order to explain the presence of such stars, we
consider various Galactic evolution mechanisms.

{It has been suggested by \citet{Haywood13} that a turbulent ISM
\citep{Brook04,Bournaud09} in the inner Galaxy could allow gas
enrichment to reach solar metallicity within a few Gyr. 
SMR stars born in the inner Galaxy would then experience interactions 
with inhomogeneities in the primitive disc to bring them to the solar neighbourhood.
However, \citet{Kordopatis15_rich} note that stars formed in the early turbulent ISM
would now be on highly eccentric orbits, which is not the case for most SMR
stars in RAVE (see their Figs. 9 and 10). It is also possible that these stars were formed in gas clouds
on non-circular orbits (e.g., from gas being accreted from outside the disc),
and therefore the stars would, from birth, be on kinematically hotter orbits
themselves. However, such gas is typically of sub-solar metallicity
\citep{Wakker01,Richter01,Richter06} and therefore would most likely not give
rise to the metal-rich population that we see.}

{Now consider the possibility that the SMR stars were born
kinematically cold. If these stars were scattered onto more
eccentric/inclined orbits by either the Lindblad
resonances of the spiral arms or by giant molecular clouds
{(`blurring')}, then these SMR stars in the {$\alpha$-low
component} would be visitors from the inner galaxy at the apocentres of their
orbits. On this account they would lag the LSR as we observe. We
find that our most metal-rich stars ($\feh\ga0.3$) lag the LSR by
$\sim10\kms$. \citet{Golubov13} argue that this asymmetric drift should also
be reflected in increasing radial velocity dispersion, i.e., increasing
$\sigma_{{R}}$ with decreasing $\langle V_{\phi}\rangle$. For our
{$\alpha$-low component}, we find the trends of
$\sigma_{{R},\phi,{Z}}$ with [Fe/H] to be flat --
Fig.~\ref{fig:vphi_met_std} shows just a hint of increasing $\sigma_{{R}}$
and $\sigma_{\phi}$.}

Although blurring may influence the {relationship between
chemistry, kinematics and position that} we find, we still need to explain
the finding of
\citet{Kordopatis15_rich}  that SMR stars in RAVE ({defined as
$\feh>0.1$}) follow roughly circular orbits, with approximately half of these
metal-rich stars having eccentricities below $e \sim 0.15$. 
On the other hand, churning involves an increase in angular
momentum and thus guiding radius without any increase in eccentricity.
{As a large fraction of these SMR stars have circular orbits, we
consider it likely that these stars have been brought to the solar
neighbourhood largely by churning.}

{A number of high-resolution studies find the SMR stars in the
$\alpha$-low component have a relatively small spread in [Mg/Fe]
\citep{Recio-Blanco14,Bensby14,Kordopatis15_gaiaeso, Haywood13}
and we find a
slight indication of this at the SMR tail of our $\alpha$-low component. This
scenario is consistent with the model presented in \citet{Nidever14},
according to which the $\alpha$-low component arises by superposition of
populations with differing star formation histories. This manifests as a
relatively narrow sequence of $\alpha$-low stars over a large [Fe/H] range.
\citet{Nidever14} note that a possible origin of this effect is described in
\citet{Schoenrich09_rm}, where the superposition of populations is caused by
radial migration of stars from the inner Galaxy with different birth radii
and enrichment histories. While they find that stars may experience both
blurring and churning, the effect of churning is stronger in the inner
regions of the galaxy, where SMR stars were most likely born.}
Therefore, we also consider that both mechanisms may be at work. If a star
is first blurred such that it is at larger Galactic radii, it is possible
that it may then experience a change in guiding radius due to an interaction
with a {varying} non-axisymmetric potential at corotation, such as
transient spiral arms \citep{Sellwood02}, or transient overdensities at the
bar-spiral interface \citep{Minchev13}. While it is possible for both
mechanisms to alter the kinematics of a given star, \citet{Minchev13} note
that stars on circular orbits are more likely to be affected by churning.

\section{Summary and conclusions}
\label{sec:conclusion}

We have explored the relationship between kinematics and {elemental
abundances} for a sample of extended solar neighbourhood stars obtained by
RAVE. Since high-resolution studies
\citep[e.g.][]{Recio-Blanco14,Haywood13,Bensby14,Kordopatis15_gaiaeso} have
shown that the trough in the $\feh-\afe$ plane between the $\alpha$-low and
$\alpha$-high components of the Galactic disc is narrower than the
uncertainties in $\afe$ in the RAVE survey, we have identified the RAVE stars
that are most and least likely to be members of the $\alpha$-low component.
Specifically, a star enters our $\alpha$-low sample if its location in the
$\feh-\afe$ plane is made ten times more probable by the hypothesis that
it belongs to the $\alpha$-low component than the hypothesis that it
belongs to the $\alpha$-high component. Conversely, the locations of our
$\alpha$-high stars are ten times more probable under the hypothesis that
they belong to the $\alpha$-high component than under the hypothesis that
they belong to the $\alpha$-low component.
{With this probabilistic separation criterion, we successfully determine separate kinematics for the $\alpha$-low and -high populations, for a conservative metallicity range where a two-component model is plausible. In addition, we find cool, thin-disc-like kinematics for the majority of our sample above solar metallicity.}

For the {$\alpha$-low sequence}, we find a negative trend in the mean
rotational velocity as a function of metallicity: $\partial
V_{\phi}/\partial\mathrm{[Fe/H]}=(-11\pm1)\kms\dex^{-1}$, which is a
shallower gradient than those measured by high-resolution studies of the
solar neighbourhood. For the $\alpha$-high component, we find a positive
correlation of mean rotational velocity with metallicity: $\partial
V_{\phi}/\partial\mathrm{[Fe/H]}=(51\pm10)\kms\dex^{-1}$, which agrees with
results from both low- and high-resolution surveys. 

Although a faint signature of this trend can be seen in the metal-rich 
bins of $\alpha$-high sequence {(open symbols in panel d of Fig.~\ref{fig:vphi_met})}, we note that 
this may be due to contamination by
{$\alpha$-low} stars arising from the large errors in [Fe/H].  Also at the
high-[Fe/H] end we notice a relative overabundance of $\alpha$-high stars. A
similar population of $\alpha$-high, metal-rich stars was detected by
\citet{Gazzano13}, who concluded that these objects probably belong to the
thin disc. {By contrast, \citet{Masseron15} consider it uncertain
whether these stars should be assigned to the thin or thick disc on the
grounds that these stars may have a variety of origins. Furthermore, it
is possible that {some of} these $\alpha$-high stars are young
\citep{Chiappini15,Martig15}, which suggests that they are part of the thin
disc {(however, see \citealt{Jofre14} for a discussion on
the existence of this population)}. For these reasons, we do not 
consider the kinematics of the SMR ($\feh\ga0.15$) stars in the $\alpha$-high
component.}

The $\alpha$-low and $\alpha$-high components follow different
trends for all three components of the velocity dispersion. While the
velocity dispersions of the chemical disc components are similar in
the metallicity regime $-0.2 < \mathrm{[Fe/H]} < 0.0$, there are significant
differences {at the metal-poor end}. The mean dispersion of a given
velocity component is $\sim16\kms$ less for {$\alpha$-low} stars than
$\alpha$-high stars.  Notwithstanding  some contamination of one component by
the other, our chemically separated {components} exhibit
markedly different kinematics, which are consistent with the trends found
using higher resolution data.

RAVE offers a unique statistical opportunities to constrain theories of
Galaxy evolution.  While high-resolution surveys will have a very small
overlap with Gaia DR1, $\sim 3\times10^{5}$ RAVE stars are expected to be in
Gaia DR1.  Hence RAVE data combined with more accurate parallaxes and proper
motions from Gaia DR1 should significantly sharpen, and hopefully confirm,
the chemodynamical trends reported here and enable us to track more securely
the extent and effect of radial migration in the Galactic discs.

\section*{Acknowledgements}

We thank the referee for their thorough comments and suggestions which have
helped improve the quality of the manuscript. We also thank Philipp Richter, Ivan Minchev, Friedrich Anders, 
Cristina Chiappini, and Else Starkenburg for their comments and helpful discussions, which have
improved the quality and clarity of the text. 
Funding for this work and for RAVE has been provided by: the
Australian Astronomical Observatory; the Leibniz-Institut
fuer Astrophysik Potsdam (AIP); the Australian National
University; the Australian Research Council; the European Research Council
under the European Union's Seventh Framework
Programme (Grant Agreement 240271 and 321067); the French National
Research Agency; the German Research Foundation
(SPP 1177 and SFB 881); the Istituto Nazionale di
Astrosica at Padova; The Johns Hopkins University; the
National Science Foundation of the USA (AST-0908326);
the W. M. Keck foundation; the Macquarie University;
the Netherlands Research School for Astronomy; the Natural
Sciences and Engineering Research Council of Canada;
the Slovenian Research Agency; the Swiss National Science
Foundation; the Science \& Technology Facilities Council of
the UK; Opticon; Strasbourg Observatory; and the Universities
of Groningen, Heidelberg and Sydney. The RAVE web
site is at https://www.rave-survey.org.




\bibliographystyle{mnras}
\bibliography{mybib} 

\begin{thebibliography}{}
\makeatletter
\relax
\def\mn@urlcharsother{\let\do\@makeother \do\$\do\&\do\#\do\^\do\_\do\%\do\~}
\def\mn@doi{\begingroup\mn@urlcharsother \@ifnextchar [ {\mn@doi@}
  {\mn@doi@[]}}
\def\mn@doi@[#1]#2{\def\@tempa{#1}\ifx\@tempa\@empty \href
  {http://dx.doi.org/#2} {doi:#2}\else \href {http://dx.doi.org/#2} {#1}\fi
  \endgroup}
\def\mn@eprint#1#2{\mn@eprint@#1:#2::\@nil}
\def\mn@eprint@arXiv#1{\href {http://arxiv.org/abs/#1} {{\tt arXiv:#1}}}
\def\mn@eprint@dblp#1{\href {http://dblp.uni-trier.de/rec/bibtex/#1.xml}
  {dblp:#1}}
\def\mn@eprint@#1:#2:#3:#4\@nil{\def\@tempa {#1}\def\@tempb {#2}\def\@tempc
  {#3}\ifx \@tempc \@empty \let \@tempc \@tempb \let \@tempb \@tempa \fi \ifx
  \@tempb \@empty \def\@tempb {arXiv}\fi \@ifundefined
  {mn@eprint@\@tempb}{\@tempb:\@tempc}{\expandafter \expandafter \csname
  mn@eprint@\@tempb\endcsname \expandafter{\@tempc}}}

\bibitem[\protect\citeauthoryear{{Adibekyan} et~al.,}{{Adibekyan}
  et~al.}{2013}]{Adibekyan13}
{Adibekyan} V.~Z.,  et~al., 2013, \mn@doi [\aap] {10.1051/0004-6361/201321520},
  \href {http://adsabs.harvard.edu/abs/2013A%26A...554A..44A} {554, A44}

\bibitem[\protect\citeauthoryear{{Aumer} \& {Binney}}{{Aumer} \&
  {Binney}}{2009}]{Aumer09}
{Aumer} M.,  {Binney} J.~J.,  2009, \mn@doi [\mnras]
  {10.1111/j.1365-2966.2009.15053.x}, \href
  {http://adsabs.harvard.edu/abs/2009MNRAS.397.1286A} {397, 1286}

\bibitem[\protect\citeauthoryear{{Aumer}, {Binney}  \& {Sch{\"o}nrich}}{{Aumer}
  et~al.}{2016}]{AumerBS}
{Aumer} M.,  {Binney} J.,   {Sch{\"o}nrich} R.,  2016, \mn@doi [\mnras]
  {10.1093/mnras/stw777}, \href
  {http://ukads.nottingham.ac.uk/abs/2016MNRAS.tmp..557A} {}

\bibitem[\protect\citeauthoryear{{Bensby}, {Feltzing}  \&
  {Lundstr{\"o}m}}{{Bensby} et~al.}{2003}]{Bensby03}
{Bensby} T.,  {Feltzing} S.,   {Lundstr{\"o}m} I.,  2003, \mn@doi [\aap]
  {10.1051/0004-6361:20031213}, \href
  {http://adsabs.harvard.edu/abs/2003A%26A...410..527B} {410, 527}

\bibitem[\protect\citeauthoryear{{Bensby}, {Feltzing}, {Lundstr{\"o}m}  \&
  {Ilyin}}{{Bensby} et~al.}{2005}]{Bensby05}
{Bensby} T.,  {Feltzing} S.,  {Lundstr{\"o}m} I.,   {Ilyin} I.,  2005, \mn@doi
  [\aap] {10.1051/0004-6361:20040332}, \href
  {http://adsabs.harvard.edu/abs/2005A%26A...433..185B} {433, 185}

\bibitem[\protect\citeauthoryear{{Bensby}, {Alves-Brito}, {Oey}, {Yong}  \&
  {Mel{\'e}ndez}}{{Bensby} et~al.}{2011}]{Bensby11}
{Bensby} T.,  {Alves-Brito} A.,  {Oey} M.~S.,  {Yong} D.,   {Mel{\'e}ndez} J.,
  2011, \mn@doi [\apjl] {10.1088/2041-8205/735/2/L46}, \href
  {http://adsabs.harvard.edu/abs/2011ApJ...735L..46B} {735, L46}

\bibitem[\protect\citeauthoryear{{Bensby}, {Feltzing}  \& {Oey}}{{Bensby}
  et~al.}{2014}]{Bensby14}
{Bensby} T.,  {Feltzing} S.,   {Oey} M.~S.,  2014, \mn@doi [\aap]
  {10.1051/0004-6361/201322631}, \href
  {http://adsabs.harvard.edu/abs/2014A%26A...562A..71B} {562, A71}

\bibitem[\protect\citeauthoryear{{Bergemann} et~al.,}{{Bergemann}
  et~al.}{2014}]{Bergemann14}
{Bergemann} M.,  et~al., 2014, \mn@doi [\aap] {10.1051/0004-6361/201423456},
  \href {http://adsabs.harvard.edu/abs/2014A%26A...565A..89B} {565, A89}

\bibitem[\protect\citeauthoryear{{Bertelli}, {Girardi}, {Marigo}  \&
  {Nasi}}{{Bertelli} et~al.}{2008}]{Bertelli08}
{Bertelli} G.,  {Girardi} L.,  {Marigo} P.,   {Nasi} E.,  2008, \mn@doi [\aap]
  {10.1051/0004-6361:20079165}, \href
  {http://adsabs.harvard.edu/abs/2008A%26A...484..815B} {484, 815}

\bibitem[\protect\citeauthoryear{{Binney} et~al.,}{{Binney}
  et~al.}{2014a}]{Binney14}
{Binney} J.,  et~al., 2014a, \mn@doi [\mnras] {10.1093/mnras/stt1896}, \href
  {http://adsabs.harvard.edu/abs/2014MNRAS.437..351B} {437, 351}

\bibitem[\protect\citeauthoryear{{Binney} et~al.,}{{Binney}
  et~al.}{2014b}]{Binney14_dynamics}
{Binney} J.,  et~al., 2014b, \mn@doi [\mnras] {10.1093/mnras/stt2367}, \href
  {http://adsabs.harvard.edu/abs/2014MNRAS.439.1231B} {439, 1231}

\bibitem[\protect\citeauthoryear{{Bland-Hawthorn} \&
  {Gerhard}}{{Bland-Hawthorn} \& {Gerhard}}{2016}]{Bland-Hawthorn16}
{Bland-Hawthorn} J.,  {Gerhard} O.,  2016, preprint, \href
  {http://adsabs.harvard.edu/abs/2016arXiv160207702B} {} (\mn@eprint {arXiv}
  {1602.07702})

\bibitem[\protect\citeauthoryear{{Boeche} et~al.,}{{Boeche}
  et~al.}{2011}]{Boeche11}
{Boeche} C.,  et~al., 2011, \mn@doi [\aj] {10.1088/0004-6256/142/6/193}, \href
  {http://adsabs.harvard.edu/abs/2011AJ....142..193B} {142, 193}

\bibitem[\protect\citeauthoryear{{Boeche} et~al.,}{{Boeche}
  et~al.}{2013}]{Boeche13b}
{Boeche} C.,  et~al., 2013, \mn@doi [\aap] {10.1051/0004-6361/201322085}, \href
  {http://adsabs.harvard.edu/abs/2013A%26A...559A..59B} {559, A59}

\bibitem[\protect\citeauthoryear{{Bournaud}, {Elmegreen}  \&
  {Martig}}{{Bournaud} et~al.}{2009}]{Bournaud09}
{Bournaud} F.,  {Elmegreen} B.~G.,   {Martig} M.,  2009, \mn@doi [\apjl]
  {10.1088/0004-637X/707/1/L1}, \href
  {http://adsabs.harvard.edu/abs/2009ApJ...707L...1B} {707, L1}

\bibitem[\protect\citeauthoryear{{Bovy}, {Rix}  \& {Hogg}}{{Bovy}
  et~al.}{2012a}]{Bovy12}
{Bovy} J.,  {Rix} H.-W.,   {Hogg} D.~W.,  2012a, \mn@doi [\apj]
  {10.1088/0004-637X/751/2/131}, \href
  {http://adsabs.harvard.edu/abs/2012ApJ...751..131B} {751, 131}

\bibitem[\protect\citeauthoryear{{Bovy}, {Rix}, {Liu}, {Hogg}, {Beers}  \&
  {Lee}}{{Bovy} et~al.}{2012b}]{Bovy12_sl}
{Bovy} J.,  {Rix} H.-W.,  {Liu} C.,  {Hogg} D.~W.,  {Beers} T.~C.,   {Lee}
  Y.~S.,  2012b, \mn@doi [\apj] {10.1088/0004-637X/753/2/148}, \href
  {http://adsabs.harvard.edu/abs/2012ApJ...753..148B} {753, 148}

\bibitem[\protect\citeauthoryear{{Bovy}, {Bird}, {Garc{\'{\i}}a P{\'e}rez},
  {Majewski}, {Nidever}  \& {Zasowski}}{{Bovy} et~al.}{2015}]{Bovy15}
{Bovy} J.,  {Bird} J.~C.,  {Garc{\'{\i}}a P{\'e}rez} A.~E.,  {Majewski} S.~R.,
  {Nidever} D.~L.,   {Zasowski} G.,  2015, \mn@doi [\apj]
  {10.1088/0004-637X/800/2/83}, \href
  {http://adsabs.harvard.edu/abs/2015ApJ...800...83B} {800, 83}

\bibitem[\protect\citeauthoryear{{Brook}, {Kawata}, {Gibson}  \&
  {Freeman}}{{Brook} et~al.}{2004}]{Brook04}
{Brook} C.~B.,  {Kawata} D.,  {Gibson} B.~K.,   {Freeman} K.~C.,  2004, \mn@doi
  [\apj] {10.1086/422709}, \href
  {http://adsabs.harvard.edu/abs/2004ApJ...612..894B} {612, 894}

\bibitem[\protect\citeauthoryear{{Cartledge}, {Lauroesch}, {Meyer}  \&
  {Sofia}}{{Cartledge} et~al.}{2006}]{Cartledge06}
{Cartledge} S.~I.~B.,  {Lauroesch} J.~T.,  {Meyer} D.~M.,   {Sofia} U.~J.,
  2006, \mn@doi [\apj] {10.1086/500297}, \href
  {http://adsabs.harvard.edu/abs/2006ApJ...641..327C} {641, 327}

\bibitem[\protect\citeauthoryear{{Chen} et~al.,}{{Chen} et~al.}{2001}]{Chen01}
{Chen} B.,  et~al., 2001, \mn@doi [\apj] {10.1086/320647}, \href
  {http://adsabs.harvard.edu/abs/2001ApJ...553..184C} {553, 184}

\bibitem[\protect\citeauthoryear{{Chiappini}, {Matteucci}  \&
  {Gratton}}{{Chiappini} et~al.}{1997}]{Chiappini97}
{Chiappini} C.,  {Matteucci} F.,   {Gratton} R.,  1997, \apj, \href
  {http://adsabs.harvard.edu/abs/1997ApJ...477..765C} {477, 765}

\bibitem[\protect\citeauthoryear{{Chiappini} et~al.,}{{Chiappini}
  et~al.}{2015}]{Chiappini15}
{Chiappini} C.,  et~al., 2015, \mn@doi [\aap] {10.1051/0004-6361/201525865},
  \href {http://adsabs.harvard.edu/abs/2015A%26A...576L..12C} {576, L12}

\bibitem[\protect\citeauthoryear{{Chiba} \& {Beers}}{{Chiba} \&
  {Beers}}{2000}]{Chiba00}
{Chiba} M.,  {Beers} T.~C.,  2000, \mn@doi [\aj] {10.1086/301409}, \href
  {http://adsabs.harvard.edu/abs/2000AJ....119.2843C} {119, 2843}

\bibitem[\protect\citeauthoryear{{Cutri} et~al.,}{{Cutri}
  et~al.}{2003}]{Cutri03}
{Cutri} R.~M.,  et~al., 2003, VizieR Online Data Catalog, \href
  {http://adsabs.harvard.edu/abs/2003yCat.2246....0C} {2246}

\bibitem[\protect\citeauthoryear{{De Silva} et~al.,}{{De Silva}
  et~al.}{2015}]{DeSilva15}
{De Silva} G.~M.,  et~al., 2015, \mn@doi [\mnras] {10.1093/mnras/stv327}, \href
  {http://adsabs.harvard.edu/abs/2015MNRAS.449.2604D} {449, 2604}

\bibitem[\protect\citeauthoryear{{Edvardsson}, {Andersen}, {Gustafsson},
  {Lambert}, {Nissen}  \& {Tomkin}}{{Edvardsson} et~al.}{1993}]{Edvardsson93}
{Edvardsson} B.,  {Andersen} J.,  {Gustafsson} B.,  {Lambert} D.~L.,  {Nissen}
  P.~E.,   {Tomkin} J.,  1993, \aap, \href
  {http://adsabs.harvard.edu/abs/1993A%26A...275..101E} {275, 101}

\bibitem[\protect\citeauthoryear{{Freeman} \& {Bland-Hawthorn}}{{Freeman} \&
  {Bland-Hawthorn}}{2002}]{Freeman02}
{Freeman} K.,  {Bland-Hawthorn} J.,  2002, \mn@doi [\araa]
  {10.1146/annurev.astro.40.060401.093840}, \href
  {http://adsabs.harvard.edu/abs/2002ARA%26A..40..487F} {40, 487}

\bibitem[\protect\citeauthoryear{{Fuhrmann}}{{Fuhrmann}}{1998}]{Fuhrmann98}
{Fuhrmann} K.,  1998, \aap, \href
  {http://adsabs.harvard.edu/abs/1998A%26A...338..161F} {338, 161}

\bibitem[\protect\citeauthoryear{{Fuhrmann}}{{Fuhrmann}}{2004}]{Fuhrmann04}
{Fuhrmann} K.,  2004, \mn@doi [Astronomische Nachrichten]
  {10.1002/asna.200310173}, \href
  {http://adsabs.harvard.edu/abs/2004AN....325....3F} {325, 3}

\bibitem[\protect\citeauthoryear{{Fuhrmann}}{{Fuhrmann}}{2008}]{Fuhrmann08}
{Fuhrmann} K.,  2008, \mn@doi [\mnras] {10.1111/j.1365-2966.2007.12671.x},
  \href {http://adsabs.harvard.edu/abs/2008MNRAS.384..173F} {384, 173}

\bibitem[\protect\citeauthoryear{{Fuhrmann}}{{Fuhrmann}}{2011}]{Fuhrmann11}
{Fuhrmann} K.,  2011, \mn@doi [\mnras] {10.1111/j.1365-2966.2011.18476.x},
  \href {http://adsabs.harvard.edu/abs/2011MNRAS.414.2893F} {414, 2893}

\bibitem[\protect\citeauthoryear{{Gazzano}, {Kordopatis}, {Deleuil}, {de
  Laverny}, {Recio-Blanco}  \& {Hill}}{{Gazzano} et~al.}{2013}]{Gazzano13}
{Gazzano} J.-C.,  {Kordopatis} G.,  {Deleuil} M.,  {de Laverny} P.,
  {Recio-Blanco} A.,   {Hill} V.,  2013, \mn@doi [\aap]
  {10.1051/0004-6361/201117747}, \href
  {http://adsabs.harvard.edu/abs/2013A%26A...550A.125G} {550, A125}

\bibitem[\protect\citeauthoryear{{Genovali} et~al.,}{{Genovali}
  et~al.}{2014}]{Genovali14}
{Genovali} K.,  et~al., 2014, \mn@doi [\aap] {10.1051/0004-6361/201323198},
  \href {http://adsabs.harvard.edu/abs/2014A%26A...566A..37G} {566, A37}

\bibitem[\protect\citeauthoryear{{Gilmore}, {Wyse}  \& {Kuijken}}{{Gilmore}
  et~al.}{1989}]{Gilmore89}
{Gilmore} G.,  {Wyse} R.~F.~G.,   {Kuijken} K.,  1989, \mn@doi [\araa]
  {10.1146/annurev.aa.27.090189.003011}, \href
  {http://adsabs.harvard.edu/abs/1989ARA%26A..27..555G} {27, 555}

\bibitem[\protect\citeauthoryear{{Gilmore} et~al.,}{{Gilmore}
  et~al.}{2012}]{Gilmore12}
{Gilmore} G.,  et~al., 2012, The Messenger, \href
  {http://adsabs.harvard.edu/abs/2012Msngr.147...25G} {147, 25}

\bibitem[\protect\citeauthoryear{{Golubov} et~al.,}{{Golubov}
  et~al.}{2013}]{Golubov13}
{Golubov} O.,  et~al., 2013, \mn@doi [\aap] {10.1051/0004-6361/201321559},
  \href {http://adsabs.harvard.edu/abs/2013A%26A...557A..92G} {557, A92}

\bibitem[\protect\citeauthoryear{{Grenon}}{{Grenon}}{1972}]{Grenon72}
{Grenon} M.,  1972, in {Cayrel de Strobel} G.,  {Delplace} A.~M.,  eds, IAU
  Colloq. 17: Age des Etoiles. p.~55

\bibitem[\protect\citeauthoryear{{Guiglion} et~al.,}{{Guiglion}
  et~al.}{2015}]{Guiglion15}
{Guiglion} G.,  et~al., 2015, \mn@doi [\aap] {10.1051/0004-6361/201525883},
  \href {http://adsabs.harvard.edu/abs/2015A%26A...583A..91G} {583, A91}

\bibitem[\protect\citeauthoryear{{Hayden} et~al.,}{{Hayden}
  et~al.}{2015}]{Hayden15}
{Hayden} M.~R.,  et~al., 2015, \mn@doi [\apj] {10.1088/0004-637X/808/2/132},
  \href {http://adsabs.harvard.edu/abs/2015ApJ...808..132H} {808, 132}

\bibitem[\protect\citeauthoryear{{Haywood}}{{Haywood}}{2008}]{Haywood08}
{Haywood} M.,  2008, \mn@doi [\mnras] {10.1111/j.1365-2966.2008.13395.x}, \href
  {http://adsabs.harvard.edu/abs/2008MNRAS.388.1175H} {388, 1175}

\bibitem[\protect\citeauthoryear{{Haywood}, {Di Matteo}, {Lehnert}, {Katz}  \&
  {G{\'o}mez}}{{Haywood} et~al.}{2013}]{Haywood13}
{Haywood} M.,  {Di Matteo} P.,  {Lehnert} M.~D.,  {Katz} D.,   {G{\'o}mez} A.,
  2013, \mn@doi [\aap] {10.1051/0004-6361/201321397}, \href
  {http://adsabs.harvard.edu/abs/2013A%26A...560A.109H} {560, A109}

\bibitem[\protect\citeauthoryear{{Haywood}, {Lehnert}, {Di Matteo}, {Snaith},
  {Schultheis}, {Katz}  \& {Gomez}}{{Haywood} et~al.}{2016}]{Haywood16}
{Haywood} M.,  {Lehnert} M.~D.,  {Di Matteo} P.,  {Snaith} O.,  {Schultheis}
  M.,  {Katz} D.,   {Gomez} A.,  2016, preprint, \href
  {http://adsabs.harvard.edu/abs/2016arXiv160103042H} {} (\mn@eprint {arXiv}
  {1601.03042})

\bibitem[\protect\citeauthoryear{{Holmberg}, {Nordstr{\"o}m}  \&
  {Andersen}}{{Holmberg} et~al.}{2007}]{Holmberg07}
{Holmberg} J.,  {Nordstr{\"o}m} B.,   {Andersen} J.,  2007, \mn@doi [\aap]
  {10.1051/0004-6361:20077221}, \href
  {http://adsabs.harvard.edu/abs/2007A%26A...475..519H} {475, 519}

\bibitem[\protect\citeauthoryear{Israelian \& Meynet}{Israelian \&
  Meynet}{2008}]{Israelian08}
Israelian G.,  Meynet G.,  eds, 2008, The Metal-Rich Universe.
Cambridge University Press, \url {http://dx.doi.org/10.1017/CBO9780511536267}

\bibitem[\protect\citeauthoryear{{Jofr{\'e}} et~al.,}{{Jofr{\'e}}
  et~al.}{2014}]{Jofre14}
{Jofr{\'e}} P.,  et~al., 2014, \mn@doi [\aap] {10.1051/0004-6361/201322440},
  \href {http://adsabs.harvard.edu/abs/2014A%26A...564A.133J} {564, A133}

\bibitem[\protect\citeauthoryear{{Juri{\'c}} et~al.,}{{Juri{\'c}}
  et~al.}{2008}]{Juric08}
{Juri{\'c}} M.,  et~al., 2008, \mn@doi [\apj] {10.1086/523619}, \href
  {http://adsabs.harvard.edu/abs/2008ApJ...673..864J} {673, 864}

\bibitem[\protect\citeauthoryear{{Just} \& {Jahrei{\ss}}}{{Just} \&
  {Jahrei{\ss}}}{2010}]{Just10}
{Just} A.,  {Jahrei{\ss}} H.,  2010, \mn@doi [\mnras]
  {10.1111/j.1365-2966.2009.15893.x}, \href
  {http://adsabs.harvard.edu/abs/2010MNRAS.402..461J} {402, 461}

\bibitem[\protect\citeauthoryear{{Kordopatis} et~al.,}{{Kordopatis}
  et~al.}{2011}]{Kordopatis11}
{Kordopatis} G.,  et~al., 2011, \mn@doi [\aap] {10.1051/0004-6361/201117373},
  \href {http://adsabs.harvard.edu/abs/2011A%26A...535A.107K} {535, A107}

\bibitem[\protect\citeauthoryear{{Kordopatis} et~al.,}{{Kordopatis}
  et~al.}{2013}]{Kordopatis13}
{Kordopatis} G.,  et~al., 2013, \mn@doi [\aj] {10.1088/0004-6256/146/5/134},
  \href {http://adsabs.harvard.edu/abs/2013AJ....146..134K} {146, 134}

\bibitem[\protect\citeauthoryear{{Kordopatis} et~al.,}{{Kordopatis}
  et~al.}{2015a}]{Kordopatis15_rich}
{Kordopatis} G.,  et~al., 2015a, \mn@doi [\mnras] {10.1093/mnras/stu2726},
  \href {http://adsabs.harvard.edu/abs/2015MNRAS.447.3526K} {447, 3526}

\bibitem[\protect\citeauthoryear{{Kordopatis} et~al.,}{{Kordopatis}
  et~al.}{2015b}]{Kordopatis15_gaiaeso}
{Kordopatis} G.,  et~al., 2015b, \mn@doi [\aap] {10.1051/0004-6361/201526258},
  \href {http://adsabs.harvard.edu/abs/2015A%26A...582A.122K} {582, A122}

\bibitem[\protect\citeauthoryear{{Lee} et~al.,}{{Lee} et~al.}{2011}]{Lee11}
{Lee} Y.~S.,  et~al., 2011, \mn@doi [\apj] {10.1088/0004-637X/738/2/187}, \href
  {http://adsabs.harvard.edu/abs/2011ApJ...738..187L} {738, 187}

\bibitem[\protect\citeauthoryear{{Lindegren} \& {Feltzing}}{{Lindegren} \&
  {Feltzing}}{2013}]{Lindegren13}
{Lindegren} L.,  {Feltzing} S.,  2013, \mn@doi [\aap]
  {10.1051/0004-6361/201321057}, \href
  {http://adsabs.harvard.edu/abs/2013A%26A...553A..94L} {553, A94}

\bibitem[\protect\citeauthoryear{{Loebman}, {Ro{\v s}kar}, {Debattista},
  {Ivezi{\'c}}, {Quinn}  \& {Wadsley}}{{Loebman} et~al.}{2011}]{Loebman11}
{Loebman} S.~R.,  {Ro{\v s}kar} R.,  {Debattista} V.~P.,  {Ivezi{\'c}} {\v Z}.,
   {Quinn} T.~R.,   {Wadsley} J.,  2011, \mn@doi [\apj]
  {10.1088/0004-637X/737/1/8}, \href
  {http://adsabs.harvard.edu/abs/2011ApJ...737....8L} {737, 8}

\bibitem[\protect\citeauthoryear{{Majewski} et~al.,}{{Majewski}
  et~al.}{2015}]{Majewski15}
{Majewski} S.~R.,  et~al., 2015, preprint, \href
  {http://adsabs.harvard.edu/abs/2015arXiv150905420M} {} (\mn@eprint {arXiv}
  {1509.05420})

\bibitem[\protect\citeauthoryear{{Martig} et~al.,}{{Martig}
  et~al.}{2015}]{Martig15}
{Martig} M.,  et~al., 2015, \mn@doi [\mnras] {10.1093/mnras/stv1071}, \href
  {http://adsabs.harvard.edu/abs/2015MNRAS.451.2230M} {451, 2230}

\bibitem[\protect\citeauthoryear{{Masseron} \& {Gilmore}}{{Masseron} \&
  {Gilmore}}{2015}]{Masseron15}
{Masseron} T.,  {Gilmore} G.,  2015, \mn@doi [\mnras] {10.1093/mnras/stv1731},
  \href {http://adsabs.harvard.edu/abs/2015MNRAS.453.1855M} {453, 1855}

\bibitem[\protect\citeauthoryear{{Matijevi{\v c}} et~al.,}{{Matijevi{\v c}}
  et~al.}{2012}]{Matijevic12}
{Matijevi{\v c}} G.,  et~al., 2012, \mn@doi [\apjs]
  {10.1088/0067-0049/200/2/14}, \href
  {http://adsabs.harvard.edu/abs/2012ApJS..200...14M} {200, 14}

\bibitem[\protect\citeauthoryear{{Minchev}, {Famaey}, {Quillen}, {Dehnen},
  {Martig}  \& {Siebert}}{{Minchev} et~al.}{2012}]{Minchev12}
{Minchev} I.,  {Famaey} B.,  {Quillen} A.~C.,  {Dehnen} W.,  {Martig} M.,
  {Siebert} A.,  2012, \mn@doi [\aap] {10.1051/0004-6361/201219714}, \href
  {http://adsabs.harvard.edu/abs/2012A%26A...548A.127M} {548, A127}

\bibitem[\protect\citeauthoryear{{Minchev}, {Chiappini}  \& {Martig}}{{Minchev}
  et~al.}{2013}]{Minchev13}
{Minchev} I.,  {Chiappini} C.,   {Martig} M.,  2013, \mn@doi [\aap]
  {10.1051/0004-6361/201220189}, \href
  {http://adsabs.harvard.edu/abs/2013A%26A...558A...9M} {558, A9}

\bibitem[\protect\citeauthoryear{{Monari}, {Helmi}, {Antoja}  \&
  {Steinmetz}}{{Monari} et~al.}{2014}]{Monari14}
{Monari} G.,  {Helmi} A.,  {Antoja} T.,   {Steinmetz} M.,  2014, \mn@doi [\aap]
  {10.1051/0004-6361/201423666}, \href
  {http://adsabs.harvard.edu/abs/2014A%26A...569A..69M} {569, A69}

\bibitem[\protect\citeauthoryear{{Nidever} et~al.,}{{Nidever}
  et~al.}{2014}]{Nidever14}
{Nidever} D.~L.,  et~al., 2014, \mn@doi [\apj] {10.1088/0004-637X/796/1/38},
  \href {http://adsabs.harvard.edu/abs/2014ApJ...796...38N} {796, 38}

\bibitem[\protect\citeauthoryear{{Nissen} \& {Schuster}}{{Nissen} \&
  {Schuster}}{2010}]{Nissen10}
{Nissen} P.~E.,  {Schuster} W.~J.,  2010, \mn@doi [\aap]
  {10.1051/0004-6361/200913877}, \href
  {http://adsabs.harvard.edu/abs/2010A%26A...511L..10N} {511, L10}

\bibitem[\protect\citeauthoryear{{Nordstr{\"o}m} et~al.,}{{Nordstr{\"o}m}
  et~al.}{2004}]{Nordstroem04}
{Nordstr{\"o}m} B.,  et~al., 2004, \mn@doi [\aap] {10.1051/0004-6361:20035959},
  \href {http://adsabs.harvard.edu/abs/2004A%26A...418..989N} {418, 989}

\bibitem[\protect\citeauthoryear{{Pasetto} et~al.,}{{Pasetto}
  et~al.}{2012}]{Pasetto12}
{Pasetto} S.,  et~al., 2012, \mn@doi [\aap] {10.1051/0004-6361/201219462},
  \href {http://adsabs.harvard.edu/abs/2012A%26A...547A..71P} {547, A71}

\bibitem[\protect\citeauthoryear{{Piffl} et~al.,}{{Piffl}
  et~al.}{2014}]{Piffl14}
{Piffl} T.,  et~al., 2014, \mn@doi [\aap] {10.1051/0004-6361/201322531}, \href
  {http://adsabs.harvard.edu/abs/2014A%26A...562A..91P} {562, A91}

\bibitem[\protect\citeauthoryear{{Quillen} \& {Garnett}}{{Quillen} \&
  {Garnett}}{2001}]{Quillen01}
{Quillen} A.~C.,  {Garnett} D.~R.,  2001, in {Funes} J.~G.,  {Corsini} E.~M.,
  eds,  Astronomical Society of the Pacific Conference Series Vol. 230, Galaxy
  Disks and Disk Galaxies. pp 87--88

\bibitem[\protect\citeauthoryear{{Recio-Blanco} et~al.,}{{Recio-Blanco}
  et~al.}{2014}]{Recio-Blanco14}
{Recio-Blanco} A.,  et~al., 2014, \mn@doi [\aap] {10.1051/0004-6361/201322944},
  \href {http://adsabs.harvard.edu/abs/2014A%26A...567A...5R} {567, A5}

\bibitem[\protect\citeauthoryear{{Reddy}, {Lambert}  \& {Allende
  Prieto}}{{Reddy} et~al.}{2006}]{Reddy06}
{Reddy} B.~E.,  {Lambert} D.~L.,   {Allende Prieto} C.,  2006, \mn@doi [\mnras]
  {10.1111/j.1365-2966.2006.10148.x}, \href
  {http://adsabs.harvard.edu/abs/2006MNRAS.367.1329R} {367, 1329}

\bibitem[\protect\citeauthoryear{{Richter}}{{Richter}}{2006}]{Richter06}
{Richter} P.,  2006, in {Roeser} S.,  ed., ~ Vol. 19, Reviews in Modern
  Astronomy. p.~31 (\mn@eprint {} {astro-ph/0602343})

\bibitem[\protect\citeauthoryear{{Richter}, {Sembach}, {Wakker}, {Savage},
  {Tripp}, {Murphy}, {Kalberla}  \& {Jenkins}}{{Richter}
  et~al.}{2001}]{Richter01}
{Richter} P.,  {Sembach} K.~R.,  {Wakker} B.~P.,  {Savage} B.~D.,  {Tripp}
  T.~M.,  {Murphy} E.~M.,  {Kalberla} P.~M.~W.,   {Jenkins} E.~B.,  2001,
  \mn@doi [\apj] {10.1086/322401}, \href
  {http://adsabs.harvard.edu/abs/2001ApJ...559..318R} {559, 318}

\bibitem[\protect\citeauthoryear{{Robin}, {Reyl{\'e}}, {Derri{\`e}re}  \&
  {Picaud}}{{Robin} et~al.}{2003}]{Robin03}
{Robin} A.~C.,  {Reyl{\'e}} C.,  {Derri{\`e}re} S.,   {Picaud} S.,  2003,
  \mn@doi [\aap] {10.1051/0004-6361:20031117}, \href
  {http://adsabs.harvard.edu/abs/2003A%26A...409..523R} {409, 523}

\bibitem[\protect\citeauthoryear{{Roeser}, {Demleitner}  \&
  {Schilbach}}{{Roeser} et~al.}{2010}]{Roeser10}
{Roeser} S.,  {Demleitner} M.,   {Schilbach} E.,  2010, \mn@doi [\aj]
  {10.1088/0004-6256/139/6/2440}, \href
  {http://adsabs.harvard.edu/abs/2010AJ....139.2440R} {139, 2440}

\bibitem[\protect\citeauthoryear{{Sch{\"o}nrich}}{{Sch{\"o}nrich}}{2012}]{Schoenrich12}
{Sch{\"o}nrich} R.,  2012, \mn@doi [\mnras] {10.1111/j.1365-2966.2012.21631.x},
  \href {http://adsabs.harvard.edu/abs/2012MNRAS.427..274S} {427, 274}

\bibitem[\protect\citeauthoryear{{Sch{\"o}nrich} \& {Binney}}{{Sch{\"o}nrich}
  \& {Binney}}{2009a}]{Schoenrich09_rm}
{Sch{\"o}nrich} R.,  {Binney} J.,  2009a, \mn@doi [\mnras]
  {10.1111/j.1365-2966.2009.14750.x}, \href
  {http://adsabs.harvard.edu/abs/2009MNRAS.396..203S} {396, 203}

\bibitem[\protect\citeauthoryear{{Sch{\"o}nrich} \& {Binney}}{{Sch{\"o}nrich}
  \& {Binney}}{2009b}]{Schoenrich09}
{Sch{\"o}nrich} R.,  {Binney} J.,  2009b, \mn@doi [\mnras]
  {10.1111/j.1365-2966.2009.15365.x}, \href
  {http://adsabs.harvard.edu/abs/2009MNRAS.399.1145S} {399, 1145}

\bibitem[\protect\citeauthoryear{{Sch{\"o}nrich}, {Binney}  \&
  {Dehnen}}{{Sch{\"o}nrich} et~al.}{2010}]{Schoenrich10}
{Sch{\"o}nrich} R.,  {Binney} J.,   {Dehnen} W.,  2010, \mn@doi [\mnras]
  {10.1111/j.1365-2966.2010.16253.x}, \href
  {http://adsabs.harvard.edu/abs/2010MNRAS.403.1829S} {403, 1829}

\bibitem[\protect\citeauthoryear{{Sellwood} \& {Binney}}{{Sellwood} \&
  {Binney}}{2002}]{Sellwood02}
{Sellwood} J.~A.,  {Binney} J.~J.,  2002, \mn@doi [\mnras]
  {10.1046/j.1365-8711.2002.05806.x}, \href
  {http://adsabs.harvard.edu/abs/2002MNRAS.336..785S} {336, 785}

\bibitem[\protect\citeauthoryear{{Siebert} et~al.,}{{Siebert}
  et~al.}{2008}]{Siebert08}
{Siebert} A.,  et~al., 2008, \mn@doi [\mnras]
  {10.1111/j.1365-2966.2008.13912.x}, \href
  {http://adsabs.harvard.edu/abs/2008MNRAS.391..793S} {391, 793}

\bibitem[\protect\citeauthoryear{{Siebert} et~al.,}{{Siebert}
  et~al.}{2011}]{Siebert11}
{Siebert} A.,  et~al., 2011, \mn@doi [\mnras]
  {10.1111/j.1365-2966.2010.18037.x}, \href
  {http://adsabs.harvard.edu/abs/2011MNRAS.412.2026S} {412, 2026}

\bibitem[\protect\citeauthoryear{{Soubiran}, {Bienaym{\'e}}  \&
  {Siebert}}{{Soubiran} et~al.}{2003}]{Soubiran03}
{Soubiran} C.,  {Bienaym{\'e}} O.,   {Siebert} A.,  2003, \mn@doi [\aap]
  {10.1051/0004-6361:20021615}, \href
  {http://adsabs.harvard.edu/abs/2003A%26A...398..141S} {398, 141}

\bibitem[\protect\citeauthoryear{{Steinmetz} et~al.,}{{Steinmetz}
  et~al.}{2006}]{Steinmetz06}
{Steinmetz} M.,  et~al., 2006, \mn@doi [\aj] {10.1086/506564}, \href
  {http://adsabs.harvard.edu/abs/2006AJ....132.1645S} {132, 1645}

\bibitem[\protect\citeauthoryear{{Vickers}, {Roeser}  \& {Grebel}}{{Vickers}
  et~al.}{2016}]{Vickers16}
{Vickers} J.~J.,  {Roeser} S.,   {Grebel} E.~K.,  2016, preprint, \href
  {http://adsabs.harvard.edu/abs/2016arXiv160208868V} {} (\mn@eprint {arXiv}
  {1602.08868})

\bibitem[\protect\citeauthoryear{{Wakker}}{{Wakker}}{2001}]{Wakker01}
{Wakker} B.~P.,  2001, \mn@doi [\apjs] {10.1086/321783}, \href
  {http://adsabs.harvard.edu/abs/2001ApJS..136..463W} {136, 463}

\bibitem[\protect\citeauthoryear{{Williams} et~al.,}{{Williams}
  et~al.}{2013}]{Williams13}
{Williams} M.~E.~K.,  et~al., 2013, \mn@doi [\mnras] {10.1093/mnras/stt1522},
  \href {http://adsabs.harvard.edu/abs/2013MNRAS.436..101W} {436, 101}

\bibitem[\protect\citeauthoryear{{Yanny} et~al.,}{{Yanny}
  et~al.}{2009}]{Yanny09}
{Yanny} B.,  et~al., 2009, \mn@doi [\aj] {10.1088/0004-6256/137/5/4377}, \href
  {http://adsabs.harvard.edu/abs/2009AJ....137.4377Y} {137, 4377}

\bibitem[\protect\citeauthoryear{{Zacharias}, {Finch}, {Girard}, {Henden},
  {Bartlett}, {Monet}  \& {Zacharias}}{{Zacharias} et~al.}{2013}]{Zacharias13}
{Zacharias} N.,  {Finch} C.~T.,  {Girard} T.~M.,  {Henden} A.,  {Bartlett}
  J.~L.,  {Monet} D.~G.,   {Zacharias} M.~I.,  2013, \mn@doi [\aj]
  {10.1088/0004-6256/145/2/44}, \href
  {http://cdsads.u-strasbg.fr/abs/2013AJ....145...44Z} {145, 44}

\bibitem[\protect\citeauthoryear{{Zhao}, {Zhao}, {Chu}, {Jing}  \&
  {Deng}}{{Zhao} et~al.}{2012}]{Zhao12}
{Zhao} G.,  {Zhao} Y.-H.,  {Chu} Y.-Q.,  {Jing} Y.-P.,   {Deng} L.-C.,  2012,
  \mn@doi [Research in Astronomy and Astrophysics]
  {10.1088/1674-4527/12/7/002}, \href
  {http://adsabs.harvard.edu/abs/2012RAA....12..723Z} {12, 723}

\bibitem[\protect\citeauthoryear{{{\v Z}erjal} et~al.,}{{{\v Z}erjal}
  et~al.}{2013}]{Zerjal13}
{{\v Z}erjal} M.,  et~al., 2013, \mn@doi [\apj] {10.1088/0004-637X/776/2/127},
  \href {http://adsabs.harvard.edu/abs/2013ApJ...776..127Z} {776, 127}

\makeatother
\end{thebibliography}







\bsp	
\label{lastpage}
\end{document}